\documentclass{article}

\usepackage{psfig}
\input epsf

\usepackage{emulateapj}
\usepackage{apjfonts}

\makeatletter
\let\@internalcite\cite
\def\cite{\def\astroncite##1##2{##1\ ##2}\@internalcite}
\def\citey{\def\astroncite##1##2{##1\ (##2)}\@internalcite}
\def\@citex[#1]#2{\if@filesw\immediate\write\@auxout{\string\citation{#2}}\fi
  \def\@citea{}\@cite{\@for\@citeb:=#2\do
    {\@citea\def\@citea{; }\@ifundefined
       {b@\@citeb}{{\bf ??}\@warning
       {Citation `\@citeb' on page \thepage \space undefined}}%
{\csname b@\@citeb\endcsname}}}{#1}}
\def\@cite#1#2{#1\if@tempswa #2\fi}
\def\@biblabel#1{}
\def\astroncite#1#2{#1\ #2}
\makeatother

\def\be{\begin{equation}}
\def\ee{\end{equation}}
\def\bea{\begin{eqnarray}}
\def\eea{\end{eqnarray}}
\def\errtwo#1#2#3{{#1}^{+ #2}_{- #3}}
\def\Tbb{T_{\rm  bb}}

\def\aproxgt{\mathrel{%
      \rlap{\raise 0.511ex \hbox{$>$}}{\lower 0.511ex \hbox{$\sim$}}}}
\def\aproxlt{\mathrel{%
      \rlap{\raise 0.511ex \hbox{$<$}}{\lower 0.511ex \hbox{$\sim$}}}}

\def\cyg{Cyg X-1 }
\def\Msun{$M_\odot$}
\def\be{\begin{equation}}
\def\ee{\end{equation}}
\def\bea{\begin{eqnarray}}
\def\eea{\end{eqnarray}}

\begin{document}

\slugcomment{Submitted to The Astrophysical Journal, April 3, 1998}

\lefthead{Nowak et al.}
\righthead{RXTE Observation of Cygnus X-1: II.}

\title{RXTE Observation of Cygnus X-1: II. Timing Analysis,}

\author{Michael A. Nowak\altaffilmark{1}, Brian A. Vaughan\altaffilmark{2},
J\"orn Wilms\altaffilmark{3}, James B. Dove\altaffilmark{1,4,5}, Mitchell
C. Begelman\altaffilmark{1,5}} 

\altaffiltext{1}{JILA, University of Colorado, Campus Box 440, Boulder,
CO~80309-0440, USA; \{mnowak, dove\}@rocinante.colorado.edu,
mitch@jila.colorado.edu}
\altaffiltext{2}{Space Radiation Laboratory, California Institute of
Technology, MC 220-47, Pasadena, CA ~91125, USA; ~brian@srl.caltech.edu}
\altaffiltext{3}{Institut f\"ur Astronomie und Astrophysik,
Abt.~Astronomie, Waldh\"auser Str. 64, D-72076 T\"ubingen, Germany;
wilms@astro.uni-tuebingen.de}
\altaffiltext{4}{also, Department of Physics and Astronomy, University of
Wyoming, Laramie, WY~82071, USA}
\altaffiltext{5}{also, Department of Astrophysical and Planetary Sciences,
University of Colorado, Boulder, CO~80309, USA}

\received{March 9, 1998}
\accepted{March 10, 1998}

\begin{abstract}

We present timing analysis for a Rossi X-ray Timing Explorer observation of
Cygnus~X-1 in its hard/low state.  This was the first RXTE observation of
Cyg~X-1 taken after it transited back to this state from its soft/high
state.  RXTE's large effective area, superior timing capabilities, and
ability to obtain long, uninterrupted observations have allowed us to
obtain measurements of the power spectral density (PSD), coherence
function, and Fourier time lags to a decade lower in frequency and half a
decade higher in frequency than typically was achieved with previous
instruments.  Notable aspects of our observations include a weak 0.005\,Hz
feature in the PSD coincident with a coherence recovery; a `hardening' of
the high-frequency PSD with increasing energy; a broad frequency range
measurement of the coherence function, revealing rollovers from unity
coherence at both low and high frequency; and an accurate determination of
the Fourier time lags over two and a half decades in frequency.  As has
been noted in previous similar observations, the time delay is
approximately proportional to $f^{-0.7}$, and at a fixed Fourier frequency
the time delay of the hard X-rays compared to the softest energy channel
tends to increase logarithmically with energy.  Curiously, the
$0.01$--$0.2$\,Hz coherence between the highest and lowest energy bands
is actually slightly greater than the coherence between the second highest
and lowest energy bands. We carefully describe all of the analysis
techniques used in this paper, and we make comparisons of the data to
general theoretical expectations.  In a companion paper, we make specific
comparisons to a Compton corona model that we have successfully used to
describe the energy spectral data from this observation.

\end{abstract}

\keywords{accretion --- black hole physics --- X-rays: binaries}

\setcounter{footnote}{0}

\section{Introduction}\label{sec:intro}

\subsection{Plan of Paper}\label{sec:plan}
Recently, we have presented the results of a spectral analysis of a 20\,ks
Rossi X-ray Timing Explorer (RXTE) observation of perhaps the most famous
of the galactic black hole candidates (GBHC), Cygnus~X--1 (\cite{dove:98a},
hereafter paper I).  The results of this analysis are summarized in
\S\ref{sec:spectra} below.  Here in this paper we focus on the timing
analysis of our RXTE observation.  Specifically, we present the \cyg
power spectral density (PSD), the coherence function, and the time delays
between soft and hard lightcurve variability. We discuss in general terms
the analysis techniques that we have used.  As the cross-spectral
techniques utilized for determining time lags and coherence are unfamiliar
to many researchers, we shall adopt a somewhat ``pedagogical tone'' at
times.  Practitioners in the field themselves have not always properly
defined error bars and noise limits (cf. \S\ref{sec:simpcoh},
\S\ref{sec:lagnoise}).  Cross spectral techniques are used increasingly in
the study of X-ray binaries, and they have benefits relative to the more
conventional cross correlation function, particularly in measurements with
large signal-to-noise ($S/N$). We hope that the discussion here is of
general use.

We begin this paper with a review of the parameters of the
Cygnus~X--1/HDE~226868 system in general (section \ref{sec:cygintro}), and
the modeling of the \cyg high energy spectrum in specific (section
\ref{sec:spectra}).  The rest of the paper is devoted to studying the
variability of our RXTE \cyg data.  In section~\ref{sec:psd} we define the
periodogram for \emph{one} lightcurve, i.e. the PSD, and describe how to
account for instrumental effects. We present our results for \cyg in five
different energy bands. Comparing the lightcurves measured simultaneously
in these different energy bands can reveal further information on the
nature of the underlying physical processes.  Section \ref{sec:coher}
describes our computations of the coherence function, while
section~\ref{sec:tlinterp} describes our measurements of the time delay
between the variability components of different energy bands.  Coherence
measures the degree of linear correlation between time series as a function
of time scale, or equivalently of Fourier frequency.  In both of these
sections, we illustrate the techniques, estimates of noise levels, etc.,
with our RXTE data for \cyg. More importantly, we apply the techniques to
the \cyg data in order to help constrain physical models of this system.
In section~\ref{sec:whatitmeans} we interpret our \cyg results in terms of
simple, physical toy models. Specific results and comparisons for the
Comptonization model of paper I are presented in a companion paper
(\cite{nowak:98b}, hereafter paper III).

\subsection{Parameters of the Cygnus X-1/HDE~226868 System}\label{sec:cygintro}

Cygnus~X-1 is one of the most firmly established persistent GBHC.
Discovered in a rocket flight in 1964 (\cite{bowyer:65a}), this object was
one of the first X-ray sources known, although its exact nature was not
clear until its optical identification with the O-star HDE~226868
(\cite{hjellming:73a}). Since then, \cyg has been the subject of
observations by almost all X-ray astronomy missions. In-depth reviews have
been published by \citey{oda:77a} and \citey{liang:84a}.
Table~\ref{tab:cygpar} lists the most important parameters of the system.

The black hole mass given in Table~\ref{tab:cygpar} is lower than the
usually quoted value of $16\pm 5$\,$M_\odot$ determined by
\citey{gies:86a}.  These authors determined the mass by using a
relationship between the rotational broadening of the absorption lines of
HDE~226868 and the radial velocity, and by assuming that these two values
depend only on the mass ratio. By modeling the light curve they found
$M=33\pm 9$\,$M_\odot$ for HDE~226868.  The black hole mass of
16\,$M_\odot$ was then determined using the mass function given in
Table~\ref{tab:cygpar}.  This fairly high value has propagated into the
literature, and is quoted in most reviews on black holes (e.g.,
\cite{tanaka:95a}). It should be noted, however, that the method used by
\citey{gies:86a} is nonstandard, and that their mass estimate for
HDE~226868 disagrees with those obtained by other methods.  Masses obtained
from photometric modulations are in the range of $16\pm 2$\,\Msun for
HDE~226868 and $8\pm 3$\,\Msun for the black hole
(\cite{balog:81a,hutchings:78a}). Masses following from a determination of
the $\log g$ value of the O-star atmosphere typically result in masses of
around $15$\,\Msun for HDE~226868 (\cite{herrero:95a,sokolov:87a,aab:84a},
and references therein). Since the work by \citey{herrero:95a} represents
the ``state of the art'' in the theory of hot stellar atmospheres including
wind loss, we present their value for the mass of HDE~226868. We
consequently use a smaller black hole mass in this paper. This mass is
still high enough, however, to point without doubt to the presence of a
black hole in the system.

A problem intimately connected with estimating the black hole mass is the
determination of the orbital inclination. The value of $35^\circ$ in
Table~\ref{tab:cygpar} is presented without a statement of its
uncertainty. Inclination measurements are intrinsically imprecise because
the inclination has to be determined from difficult optical polarization
measurements and consequently the values found for the inclination are
scattered from $26^\circ$ to $67^\circ$
(\cite{bochkarev:86a,ninkov:87b,ninkov:87a,dolan:92a}, and references
therein) with most authors deriving rather lower inclinations. For these
reasons we use $i=35^\circ$, the value adopted by \citey{herrero:95a} in
their mass determination shown in Table~\ref{tab:cygpar}.

\end{multicols}
\begin{table}
\caption{Parameters for the system Cyg~X-1/HDE~226868. Values in
  parentheses are uncertainties in units of the last digit.}
\bigskip
\hfill\hbox{
\begin{tabular}{lllll} 
\hline
\hline
                       & Optical          & Compact     & System          & References \\
                       & Companion        & Object      & Parameters      &            \\
                       & HDE~226868       & \cyg        &                 &            \\
\hline
Position:              &                  &             &                 &\\
J2000.0: \hfill $\alpha$ $\quad$  &        &  & $19^{\rm h}58^{\rm m}21\fs700$ \hfill & \hfill 1~\\
         \hfill $\delta$ $\quad$  &        &  & $+35^{\circ}12'05\farcs 82$\hfill & \hfill 1~\\

galactic: \hfill $l_{{\rm II}}$ $\quad$ &  &  & $71\fdg 33$ \hfill & \hfill 1~\\
          \hfill $b_{{\rm II}}$ $\quad$ &  &  & $+3\fdg 07$ \hfill & \hfill 1~\\
\hline
Spectral Type          & O9.7\,Iab        &             &                 & \hfill 2~\\
$T_{\rm eff}$ [K]      & 32\,000       	  &             &                 & \hfill 3~\\
$E_{\rm B-V}$          & 0.95(7)          &             &                 & \hfill 4~\\
$N_{\rm H}$ [$\rm cm^{-2}$]&              &             & $6(2)\times 10^{21}$  & \hfill 4,5~\\ 
Distance [\,kpc]       &               &             & 2.5(3)          & \hfill 6~\\
$m_{\rm V}$ [mag]      & $8.84$           &             &                 & \hfill 7~\\
$\rm B_T$ [mag]        & $9.828(22)$      &             &                 & \hfill 7~\\
$\rm V_T$ [mag]        & $9.020(17)$      &             &                 & \hfill 7~\\
$\rm B-V$ [mag]        & $+0.81$          &             &                 & \hfill 8~\\
$\rm U-B$ [mag]        & $-0.28$          &             &                 & \hfill 8~\\
Luminosity  [$\rm L_\odot$] & $10^{5.4}$  &  $10^{4}$   &                 & \hfill 3,9~\\  
Luminosity  [${\rm erg}$/s]& $10^{39}$    & $4\cdot10^{37}$   &           &  \\  
Inclination            &                  &             & 35$^\circ$      & \hfill 10~\\
$a\sin i$ [km]         &                  & $5.82(8)\times 10^6$ &        & \hfill 11~\\
Orbital Period [d]     &                  &             & 5.59974(8)      & \hfill 11~\\
Major Axis [$\rm R_\odot$] & 14.6         & 26.3        &                 &\hfill 10~\\
$v\sin i$ [km/s]       & 75.6(10)         &             &                 & \hfill 11~\\
Mass Function [$\rm M_\odot$] & 0.252(10) &             &                 & \hfill 11~\\
Mass  [$\rm M_\odot$]  & 18               & 10          &                 & \hfill 10~\\
Radius                 & 17\,$\rm R_\odot$    & 30\,km$^\dagger$ &            & \hfill 3~\\
Separation [$\rm R_\odot$]     &          &             & 41              & \hfill 10~\\
Mass loss rate [$\rm M_\odot/a$] &        &             & $3\times 10^{-6}$  & \hfill 3~\\
Wind velocity [km/s]   & 2100             &             &                 & \hfill 10~\\
\hline 
\end{tabular} }\hfill
\vspace*{3mm}

\begin{parbox}{\textwidth}
\small
\noindent
${}^1$\citey{turon:92a},
${}^2$\citey{walborn:73a},
${}^3$\citey{herrero:95a},
${}^4$\citey{wu:82a},
${}^5$\citey{balu:95a},
${}^6$\citey{ninkov:87a},
${}^7$\citey{esa:97a}, Volume~9,
${}^8$\citey{lutz:72a},
${}^9$\citey{liang:84a},
${}^{10}$see text for further
explanation,
${}^{11}$\citey{gies:82a}

\noindent${}^\dagger$ Schwarzschild radius

\noindent{Table prepared in collaboration with Katja Pottschmidt
  (\cite{pottschmidt:97a}).}

\end{parbox}


\label{tab:cygpar} 
\end{table}
\section*{}
\vskip -15pt

For our purposes, the main parameters of interest in Table~\ref{tab:cygpar}
are the compact object mass, system luminosity, and system distance.
However, few reviews have included all the potential parameters of
interest.  Furthermore, many reviews quote secondary sources.  For these
reasons, we present Table~\ref{tab:cygpar}.

\subsection{Spectral Results}\label{sec:spectra}
The X-ray spectrum of Cyg~X-1 and other galactic black hole candidates in
the hard state can be represented as the sum of a soft thermal component
and a hard, non-thermal, power law component with a high energy cutoff at
about 150\,keV.  The spectrum is further modified by reprocessing features,
in particular a weak iron fluorescence line at 6.4\,keV and possibly a
Compton reflection hump at energies around 30\,keV.  All four components
are usually explained in the context of accretion disk corona models
(\cite{hua:95a,svensson:94a,coppi:92a,st80,sunyaev:79a}). These models
assume that the thermal radiation originates in an accretion disk and is
Comptonized in a hot plasma, the ``accretion disk corona'' (ADC).
Comptonization results in a non-thermal power-law spectrum. The iron line
and the reflection hump suggest that part of the hard radiation from the
ADC is reprocessed by the comparatively cold (150\,eV) accretion disk.

The exact geometrical configuration of the accretion disk and the corona is
unknown, and is one of the major subjects of black hole X-ray astronomical
research. Due to the similarity between the hard state X-ray spectrum of
galactic black hole candidates and active galactic nuclei (AGN), most
previous work has assumed a ``sandwich configuration'' where the corona is
situated above and below the accretion disk
(\cite{poutanen:97a,hua:95a,haardt:93a} and references therein). The
evidence for a sandwich configuration in GBHC, however, is not as strong as
in the case of AGN. In previous work, some of us have pointed out that the
efficiency of Compton cooling might prevent the sandwich corona from
getting hot enough to generate the hard X-ray spectrum observed in the hard
state of Cyg~X-1
(\cite{dove:97a,wilms:97a,gierlinski:97a,poutanen:97b}). Due to the high
geometrical covering factor of the corona, about 50 per cent of all hard
radiation from the corona enters the cold accretion disk, where it is
either reflected or thermalized. The soft photons produced this way again
enter the corona where they efficiently Compton cool the plasma.  For a
given Thomson optical depth of the corona, there thus exists a maximum
coronal temperature above which no self consistent solutions exist.

Since the major physical cause preventing the sandwich configuration from
explaining the observed hard spectra is the very efficient Compton cooling,
we have sought alternative geometries that do not suffer from a large
covering factor.  The need for a smaller covering factor is also seen from
the weakness of the reflection features in GBHC compared to those in AGN.
In our previous work we have advocated one such configuration, the
``sphere+disk'' geometry (\cite{dove:97b}, see also
\cite{gierlinski:97a,poutanen:97b}).  Here, a hot spherical corona
surrounds the black hole, while the cold accretion disk which provides the
seed photons is situated farther outside. This geometry obviously has a
smaller covering factor, and in fact much higher coronal temperatures are
possible (\cite{dove:97b}).  Using self consistent numerical models
computed for the sphere+disk geometry we were able to describe successfully
the hard-state spectrum of Cyg~X-1 measured by RXTE over a broad range in
energy, $3$--$200$\,keV (\cite{dove:97b}; paper~I).  We derived an optical
depth for the spherical corona of $\tau=1.6\pm 0.1$ and an average
temperature of $kT=87\pm 5$\,keV (paper~I).

To facilitate comparison with previous work, and to test for the presence
of features in the broad-band spectrum, we also performed purely
phenomenological fits to Cyg~X-1.  We found that the 3 to 200\,keV RXTE
spectrum could be described by an exponentially cut-off power law with a
photon index $\Gamma = \errtwo{1.45}{0.01}{0.02}$, e-folding energy $E_{\rm
  f} = \errtwo{162}{9}{8}$\,keV, plus a deviation from a power law that
formally can be modeled as a thermal blackbody with temperature $k\Tbb
=\errtwo{1.2}{0.0}{0.1}$\,keV. Note that the above power law index is
harder than that typically found, which might be an indication that the
source was not in its usual ``hard state'' after exiting the ``soft state''
of 1996 (\cite{cui:96a,zhang:97b} and references therein), which ended only
a few weeks prior to our observation. No hardening was seen in the HEXTE
data from 30 to 100\,keV, indicating that any reflection features were
formally very weak. [That is, the presence of reflection needs to be
inferred from more sophisticated continuum models
(\cite{dove:98a,gierlinski:97a,poutanen:97b}), rather than from simple
  exponentially cutoff power laws reflected from cold slabs.]

\section{Power Spectral Density}\label{sec:psd}

\subsection{Techniques and Noise Levels}\label{sec:psdtech}

The RXTE observations presented here were performed October, 22, 1996. For
the timing analysis we used data obtained with the Proportional Counter
Array (PCA), the low energy instrument onboard RXTE. The PCA consists of
five nearly identical Xe proportional counters units (PCU) with a total
effective area of about 6500\,cm$^2$ (\cite{jahoda:96b}). The data
extraction was performed using the RXTE standard data analysis software,
ftools 4.0. To avoid contamination of the data due to the Earth's X-ray
bright limb, only data measured at source elevations more than $10^\circ$
above the spacecraft horizon were used.

As discussed in paper I, throughout our observations \cyg showed count
rates of $\approx 4300~{\rm cts~s}^{-1}$.  The total integrated on source
time for our observation was slightly more than 20\,ks. For approximately
2\,ks two PCU of the PCA were not operational, so these data were ignored.
We therefore had approximately 18\,ks of high quality timing
data\footnote{Note that this is slightly less than twice the duration of
the observations we used for the spectral analysis of paper I.
Approximately half the {\it standard2f} data were not included in the first
data tape that we received; however, all the timing data were included.}.
The PCA data modes for our observation are listed in Table~\ref{tab:mode},
the formal characteristics of the modes can be found in Appendix~F to
the NASA Research Announcement for RXTE (\cite{nasa:97a}).

We divided our data into 5 energy bands with ${\cal O}(700$--$1000~{\rm
cts~s}^{-1})$ each, binned at $2^{-9}$\,s resolution\footnote{We also
created lightcurves from one quarter of our $2^{-12}$\,s resolution data to
search for high-frequency signatures.  No variability in excess of counting
noise was observed above $\approx 200$\,Hz. We therefore chose to perform
our analyses with $2^{-9}$\,s resolution data.}.  We also created a very
high energy band which our energy spectral analysis indicated should be
entirely dominated by background counts. This was done to assess the level
at which background fluctuations could affect our results at lower
energies.  Thus, we created lightcurves for six energy bands in all,
labeled $1$--$6$.  The properties of these energy channels are listed in
Table~\ref{tab:chan}.

We calculated Power Spectral Densities (PSD) in each channel by dividing
the data into contiguous segments of uniform length and time resolution,
and performing a Fast Fourier Transform (FFT) on each data
segment. (Segments with data gaps of \emph{any} duration were ignored.) We
performed all the FFTs using the reverse discrete transform from IDL
(Version 4.0.1), which corresponds to
\begin{equation}
S_j ~\equiv~ \sum_{k=0}^{N-1}~s_k~ \exp(2\pi i jk/N) ~~.
\label{eq:fftdef}
\end{equation}
Here $s_k$ is the counts in the $k^{th}$ bin of the lightcurve (consisting
of $N$ evenly spaced time bins), and $S_j$ is the complex discrete Fourier
transform corresponding to the Fourier frequency $f_j = j/N ~T_s^{-1}$,
where $T_s$ is the length (in seconds) of each data segment and
$j\in[-N/2,N/2]$. From $S_j$, we form the discrete PSD,
\begin{equation}
\langle |S_j|^2 \rangle ~=~ \langle S_j^* S_j \rangle ~~,
\label{eq:psddef}
\end{equation}
where the angle brackets indicate an average over data segments
{\it and}, in most cases, over frequency bins.  Throughout we have
used a logarithmic frequency binning wherein we averaged over
frequencies $f \rightarrow f+df$, with $df/f = 0.15$.

We have employed the PSD normalization of \citey{miyamoto:91a}. In this
normalization, one multiplies $\langle |S_j|^2 \rangle $ by $2{\cal
R}^{-2}T^{-1}$, where ${\cal R}$ is the mean count rate (in ${\rm
cts~s}^{-1}$) for the energy band of interest.  As $|S_j|^2 = |S_{-j}|^2$,
we have adopted a `one sided' normalization.  The factor of 2 in the
definition of the PSD is introduced so that integrating the PSD over
positive Fourier frequencies and then taking the square root yields the
normalized root mean square (rms) variability
\begin{equation}
\mbox{rms}=\left({{\langle s^2 \rangle ~-~ \langle s
\rangle^2}\over{\langle s \rangle^2}}\right)^{1/2}~,
\end{equation}
where here the average is over individual data segments and time bins. Note
that rms variability is often quoted for the PSD integrated over restricted
frequency ranges.  We shall always explicitly state the frequency ranges
used in our measurements.  Throughout the rest of the paper we shall take
$P_s(f_j) = 2{\cal R}^{-2}T^{-1} |S(f_j)|^2$ to signify a 
PSD as calculated from a discrete FFT.  When considering two energy
channels, $P_s$ will indicate the PSD of the soft channel, and $P_h$ the
PSD of the hard channel.

To leading order (cf. \cite{leahy:83a,vanderklis:89b}), counting noise
leads to an additive ``white noise'' component.  Its PSD, which we write
$P_N(f)$, has an amplitude of $2/{\cal R}$, independent of Fourier
frequency.  Its error is ${\cal O}(1)$ for a {\it single} Fourier frequency
measured from a {\it single} data segment, and is distributed as $1/{\cal
R}$ times $\chi^2$ with 2 degrees of freedom\footnote{The fractional
uncertainty in the signal component of the PSD is also of ${\cal O}(1)$, as
measured for an individual Fourier frequency from the FFT of a single data
segment (cf. \cite{bendat,vanderklis:89b}).}.  Instrumental deadtime
(cf. \cite{zhangw:95a}) modifies this slightly.  By deadtime we mean any
process where the detection of an event subsequently changes the
sensitivity of the detectors [see \citey{zhangw:95a,zhangw:96a} for a
discussion of idealized models that can be used to approximate the effects
seen in RXTE data].  Here we use the deadtime model of \citey{zhangw:95a},
as was successfully applied by \citey{morgan:97a} to RXTE data of
{GRS~1915+105}.  The expectation value for the PSD of the noise level is
different for each energy band, and is given by
\begin{eqnarray}
P_N(f) &= &~~{{2}\over{{\cal R}_e}} \Bigg ( ~ \left [ 1 ~-~ 2 {\cal R}_{pe} 
     \tau_d ~ \left ( 1 ~-~ {{\tau_d}\over{2 t_b}} \right ) \right ]
     \nonumber \\
 & & -~  {{N_f-1}\over{N_f}} {\cal R}_{pe} \tau_d ~
     \left ( {{\tau_d}\over{t_b}} \right ) ~ \cos \left ( 2 \pi t_b f \right
     ) 
     \nonumber \\
 & &  + ~{\cal R}_{pe} {\cal R}_{vle} ~ 
      \left [ {{\sin \left( \pi \tau_{vle} f \right )} \over {\pi f}}
      \right ]^2 \Bigg ) ~~,
\label{eq:dead}
\end{eqnarray}

\end{multicols}
\begin{table}
\caption{Data modes, with their energy ranges, number of energy channels,
and time resolution (always constrained to be a power of $2$\,s by the
onboard data processors), for our \cyg observation.}
\medskip
\hfill\hbox{
\begin{tabular}{llrl}
\hline
\hline
Data Mode & Energy Range (keV) & No. Chan. & Time Res. (s) \\
\hline
B\_2ms\_8B\_0\_35\_Q  & $\quad 0$--$14.1$    & $8 \quad$  & $\quad 2^{-9}$ \\
SB\_250us\_0\_13\_2s  & $\quad 0$--$5$       & $1 \quad$  & $\quad 2^{-12}$ \\
SB\_250us\_14\_35\_2s & $\quad 5$--$14.1$    & $1 \quad$  & $\quad 2^{-12}$ \\
E\_125us\_64M\_36\_1s & $\quad 14.1$--$100$  & $64 \quad$ & $\quad 2^{-13}$ \\
\hline
\end{tabular} }\hfill
\vspace*{3mm}

\label{tab:mode} 
\caption{The energy bands used in this work (labeled $1$--$6$).  Also
listed are the corresponding PHA channels, their approximate energy range,
the mean energy of the energy band for our specific observations, and the
average count rate in the band.}
\medskip
\hfill\hbox{
\begin{tabular}{lllrr}
\hline
\hline
Energy Band & PHA Chan. & Energy Range (keV) & Mean Energy (keV) & Mean Count Rate ($\rm cts~s^{-1}$) \\
\hline
$\quad 1$ & $~0$--$10$    & $\quad 0$--$3.9$            & $2.5\qquad$   & $780\qquad$ \\
$\quad 2$ & $~11$--$16$   & $\quad 3.9$--$6.0$          & $4.8\qquad$   & $1040\qquad$ \\
$\quad 3$ & $~17$--$22$   & $\quad 6.0$--$8.2$          & $7.1\qquad$   & $730\qquad$ \\
$\quad 4$ & $~23$--$35$   & $\quad 8.2$--$14.1$         & $10.4\qquad$  & $890\qquad$ \\
$\quad 5$ & $~36$--$193$  & $\quad 14.1$--$45$          & $20.1\qquad$  & $780\qquad$ \\
$\quad 6^\dagger$ & $~194$--$255$ & $\quad 45$--$100$ &         & $70\qquad$ \\
\hline
\end{tabular} }\hfill
\begin{parbox}{\textwidth}
\small

\noindent
${}^\dagger$ Dominated by background counts.  Not used for analysis.
Background accounts for $\approx 1\%$ of the counts in band 1, up to
$\approx 8\%$ of the counts in band 5.  Therefore, none of the conclusions
in this paper should be strongly influenced by the background.
\end{parbox}
\vspace*{3mm}

\label{tab:chan}
\end{table}
\section*{}
\vskip -15pt

\noindent where ${\cal R}_e$ is the total count rate in the energy band of
interest, ${\cal R}_{pe}$ is the count rate {\it per PCU} in the energy
band of interest (as all five PCU were always on for our selected
observations, ${\cal R}_{pe} = {\cal R}_e/5$), ${\cal R}_{vle}$ is the
`very large event rate' (mainly due to high energy charged particles, and
$< 100~{\rm cts~s}^{-1}$), $\tau_d$ is the detector deadtime (for the PCA
we take $\tau_d = 10 ~\mu{\rm s}$; W. Zhang, Private Communication),
$\tau_{vle}$ is the deadtime for very large events (preset to $70~\mu{\rm
  s}$ for our observation), $N_f$ is the number of frequency bins in the
FFTs ($N_f=N/2$), and $t_b$ is the duration of the time bins in the input
lightcurves.  In practice, the terms involving the cosine and sine above
are negligible for the count rates and frequencies of interest to us.  The
above formula was compared to the $600-2048$\,Hz PSDs calculated from our
single bit data.  Assuming that those frequencies were entirely dominated
by counting noise modified by deadtime effects, eq.~(\ref{eq:dead}) was
accurate to better than $2 \times 10^{-3}/{\cal R}_e$ (absolute). Thus
eq.~(\ref{eq:dead}) correctly predicts the deadtime \emph{deviation} from
the uncorrected noise level of $2/{\cal R}_e$ to better than 25\%.

We use eq.~(\ref{eq:dead}) throughout this work to estimate the
deadtime-corrected Poisson noise level.  All averaged PSD we present have
had this level subtracted off, as have all quoted rms variability levels.
The uncertainty in the mean of the PSD is computed by dividing the {\it
  uncorrected} mean PSD by the square root of the total number of
independent segments and/or frequency bins averaged together (cf.
\cite{leahy:83a,vanderklis:89b}).  The \emph{minimum} mean PSD level
detectable with our observations is approximately the counting noise level
from eq.~(\ref{eq:dead}) divided by the square root of the number of data
segments and/or frequency bins averaged over. We explicitly show this
`effective noise level' for all of our average PSDs.  The logarithmic
binning in frequency results in an approximately $f^{-0.5}$ dependence for
the effective noise level at high frequency.  In the next section, we
describe our results for \cyg.

\subsection{Results for Cygnus X-1}\label{psd:res}

\begin{figure}
\centerline{
\psfig{figure=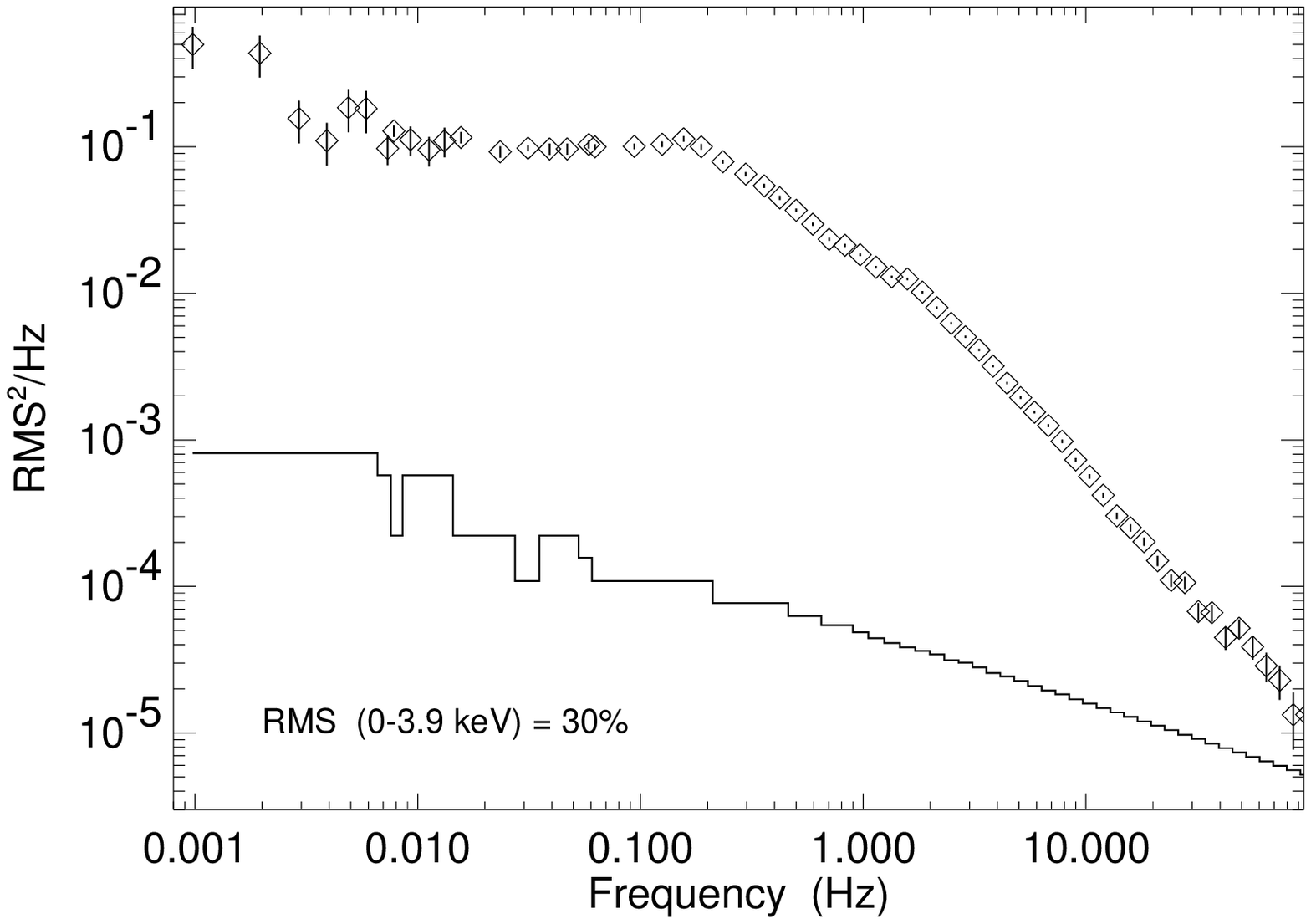,width=0.456\textwidth}
\psfig{figure=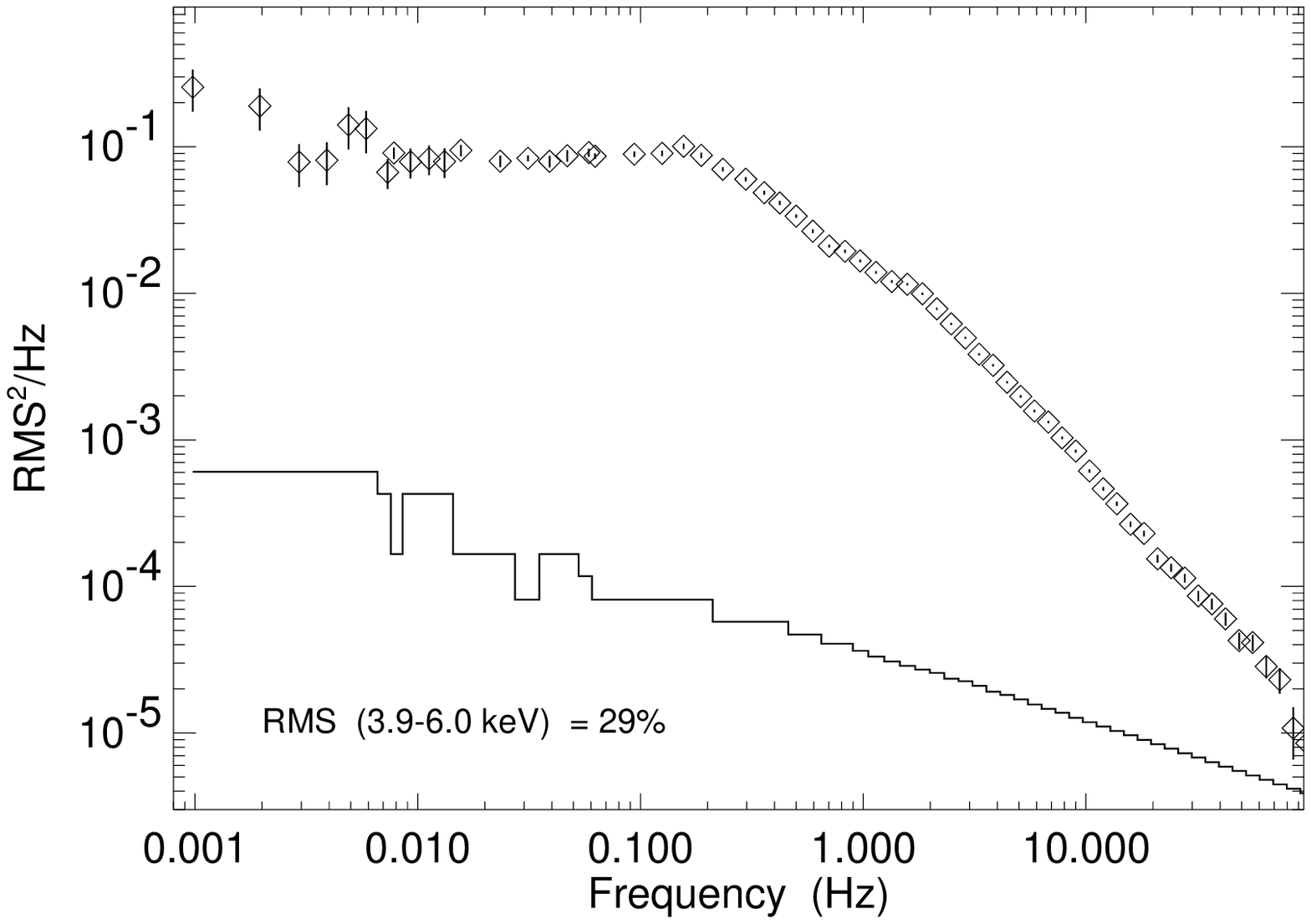,width=0.456\textwidth}
}
\centerline{
\psfig{figure=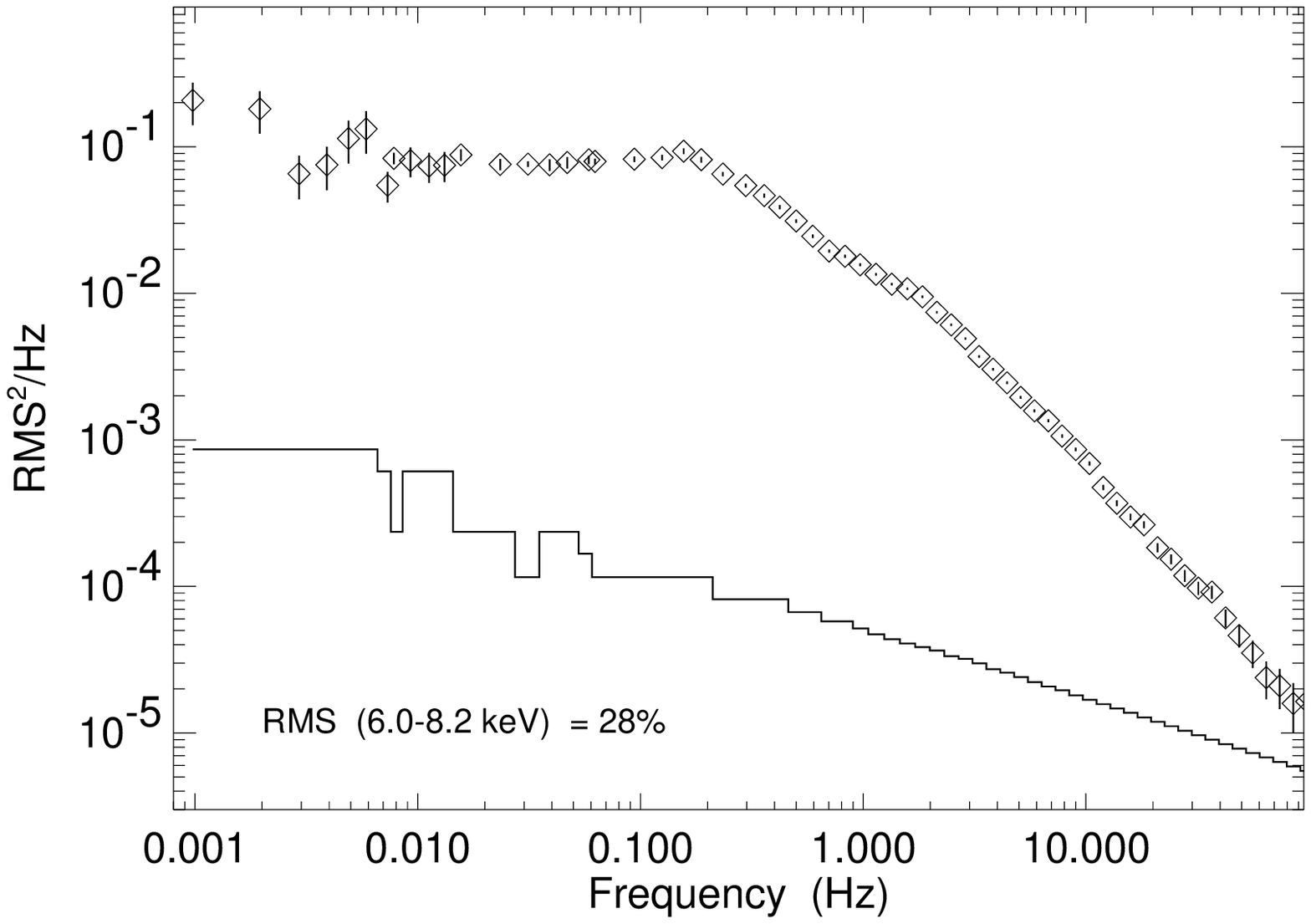,width=0.456\textwidth}
\psfig{figure=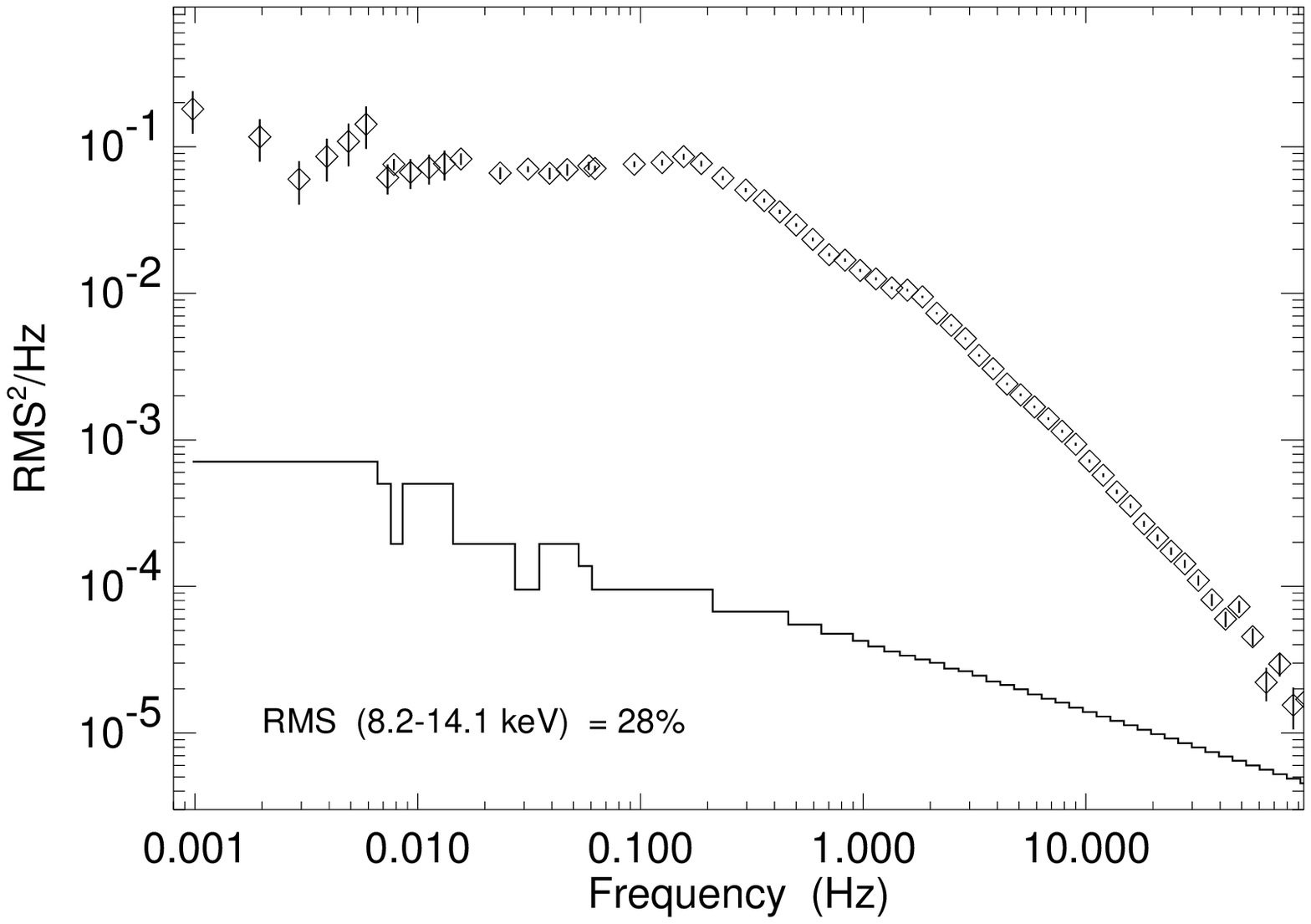,width=0.456\textwidth}
}
\centerline{
\psfig{figure=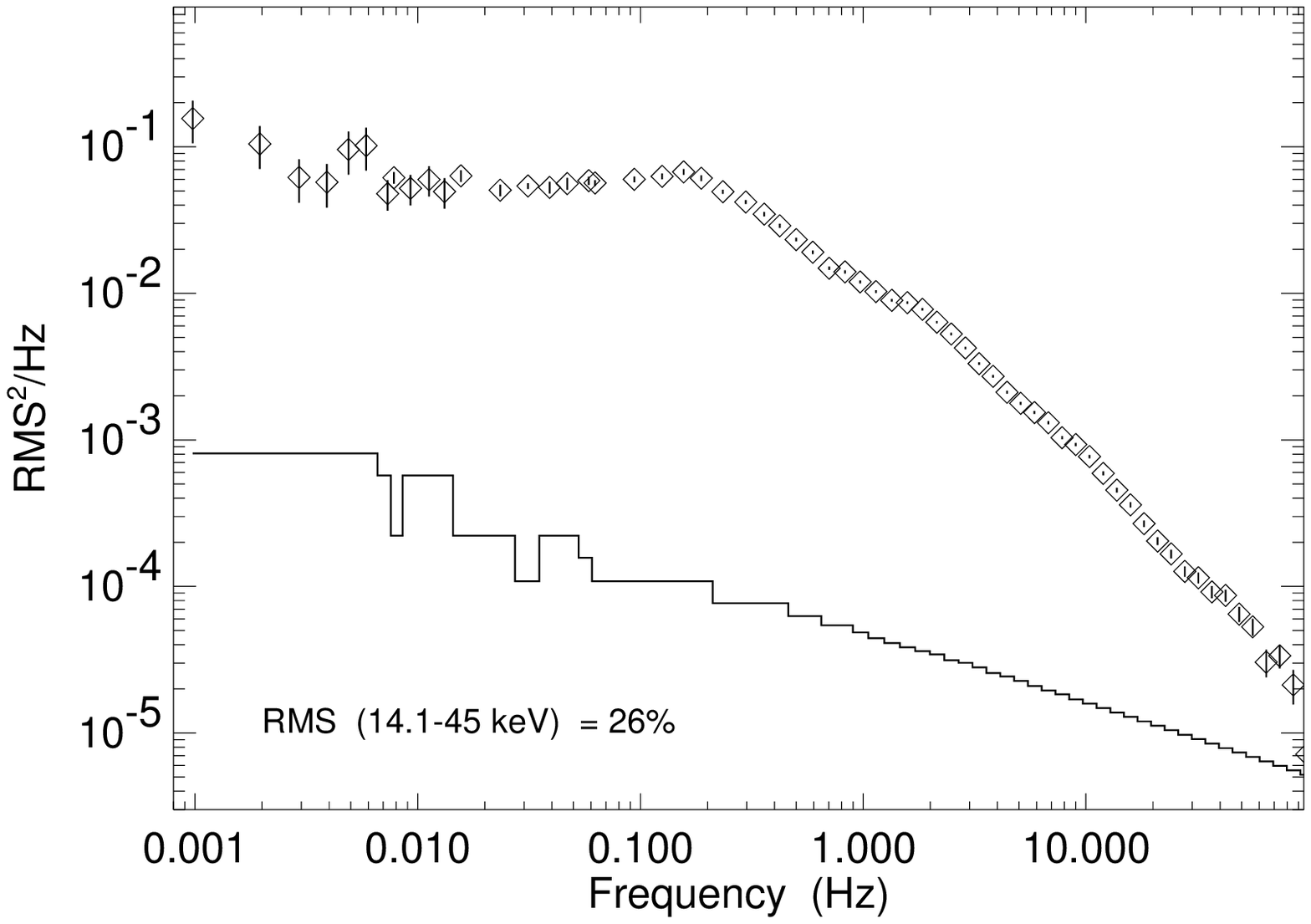,width=0.456\textwidth}
\psfig{figure=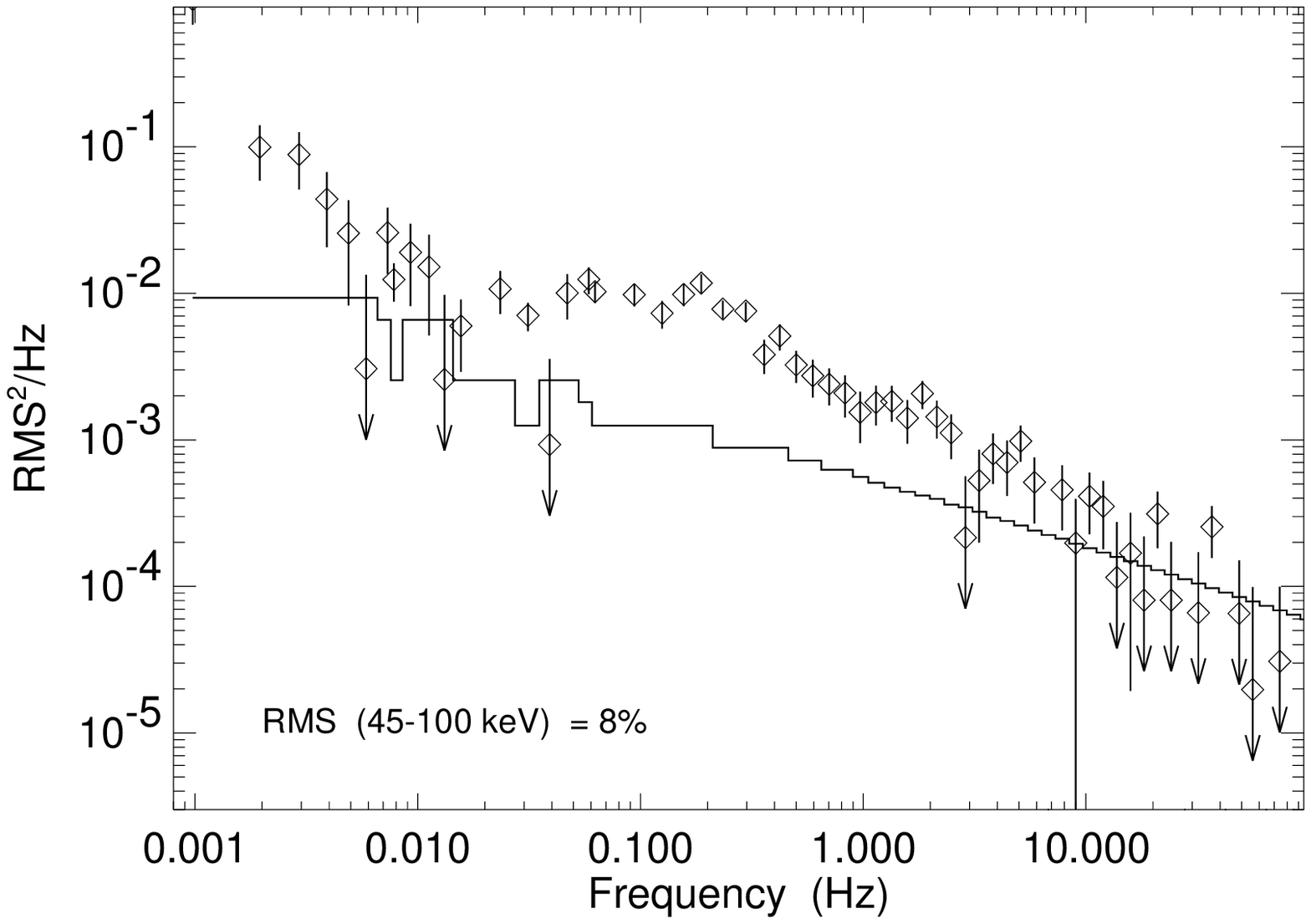,width=0.456\textwidth}
}
\caption
{Noise subtracted mean Power Spectral Densities (PSD) for the five energy
bands (and one `background' band, see text) considered in this
work. {\it Top left:} ($0-3.9$\,keV). {\it Top right:}
($3.9-6.0$\,keV). {\it Middle left:} ($6.0-8.2$\,keV). {\it Middle right:}
($8.2-14.1$\,keV). {\it Bottom left:} ($14.1-45$\,keV).  {\it Bottom
right:} ($45-100$\,keV).  For all PSDs, diamonds represent the average PSD
with a normalization such that the square root of the integrated PSD is
the fractional root mean square (rms) variability (see text and Miyamoto et
al. 1991).  The solid line represents the ``effective noise'' level, which
is essentially the minimum level of variability that could have been
detected with our observations.  All quoted rms variability levels are for
the frequency range $9.7 \times 10^{-4} - 100$\,Hz, except for the
($45-100$\,keV) band which is the $9.7 \times 10^{-4} - 10$\,Hz range.}
\label{fig:psds}
\end{figure}

We performed FFTs with three data segment lengths: 1024\,s (10 data
segments), 128\,s (133 data segments), and 32\,s (556 data segments). In
Figure~\ref{fig:psds} we show PSDs calculated from these FFTs.  As the PSDs
for each segmentation of the data are statistically correlated, the
frequency range $9.7\times10^{-4}$--$1.3\times10^{-2}$\,Hz is computed from
the 1024\,s data segments, the frequency range
$7.8\times10^{-3}$--$5.9\times10^{-2}$\,Hz is computed from the 128\,s data
segments, and the frequency range $3.1\times10^{-2}$--$256$\,Hz is computed
from the 32\,s data segments.  We use the exact same data for calculating
the coherence function and time lags described below.

The RXTE deadtime is energy dependent (\cite{zhangw:96a}), and in fact the
PSD for the background dominated, ($45$--$100$\,keV) band is above the
effective noise level.  Either eq.~(\ref{eq:dead}) is inadequate or part of
the PSD in this channel may be caused by variability of the background or
source.  However, if the 10$~\mu{\rm s}$ deadtime estimation truly is good
to better than $\approx 25\%$ at low energy, then our noise-subtracted PSDs
are reliable to at least 100 Hz.

For all energy bands, the rms variability in the $10^{-3}$--$100$\,Hz
frequency range is $\approx 30\%$.  The PSDs are relatively flat from
$0.02$--$0.2$\,Hz, then fall off at higher frequencies.  A portion of the
higher frequency range has an $f^{-1}$ dependence.  This behavior is
characteristic of the low state of BHC in general
(\cite{belloni:90a,belloni:90b,miyamoto:92a}).  There is evidence that the
PSD steepens again at frequencies below $3\times 10^{-3}$\,Hz, especially
in the lower energy bands. This low frequency noise is partly
responsible for the greater total rms in the lower energy bands. Low
frequency noise reported previously has been studied by
\citey{pottschmidt:98a} and \citey{angelini:94a}.  There also is a weak
bump, seen in all energy bands, in at least two frequency bins near
$0.005$\,Hz.  The rms variability of this region in each PSD is $\approx
1.5\%$.  This feature taken on its own is too weak to be called a
quasi-periodic oscillation (QPO).  A Lomb-Scargle periodogram
(\cite{lomb:76a,scargle:82a}) of the entire data set also shows this
feature; however, its significance is only $50$--$85\%$.  (The significance
is greatest in the lowest energy band.)  As we shall discuss in
\S\ref{sec:coher} below, however, this region of the PSD, unlike the
surrounding frequency bins, has near unity coherence. Low frequency `peaks'
in the \cyg PSD previously have been observed at frequencies ranging from
$0.04$--$0.07$\,Hz
(\cite{angelini:92a,kouveliotou:92a,angelini:94a,ubertini:94a,vikhlinin:94a}).
For the most part these features did not appear as broad peaks with a
Lorentzian profile.  Rather, they appeared as regions in the PSD where the
power law slope, as a function of Fourier frequency, discontinuously
changed from positive to negative.  Furthermore, previous observations did
not consider the coherence of these low frequency features between
different energy bands.

\begin{figure}
\centerline{
\psfig{figure=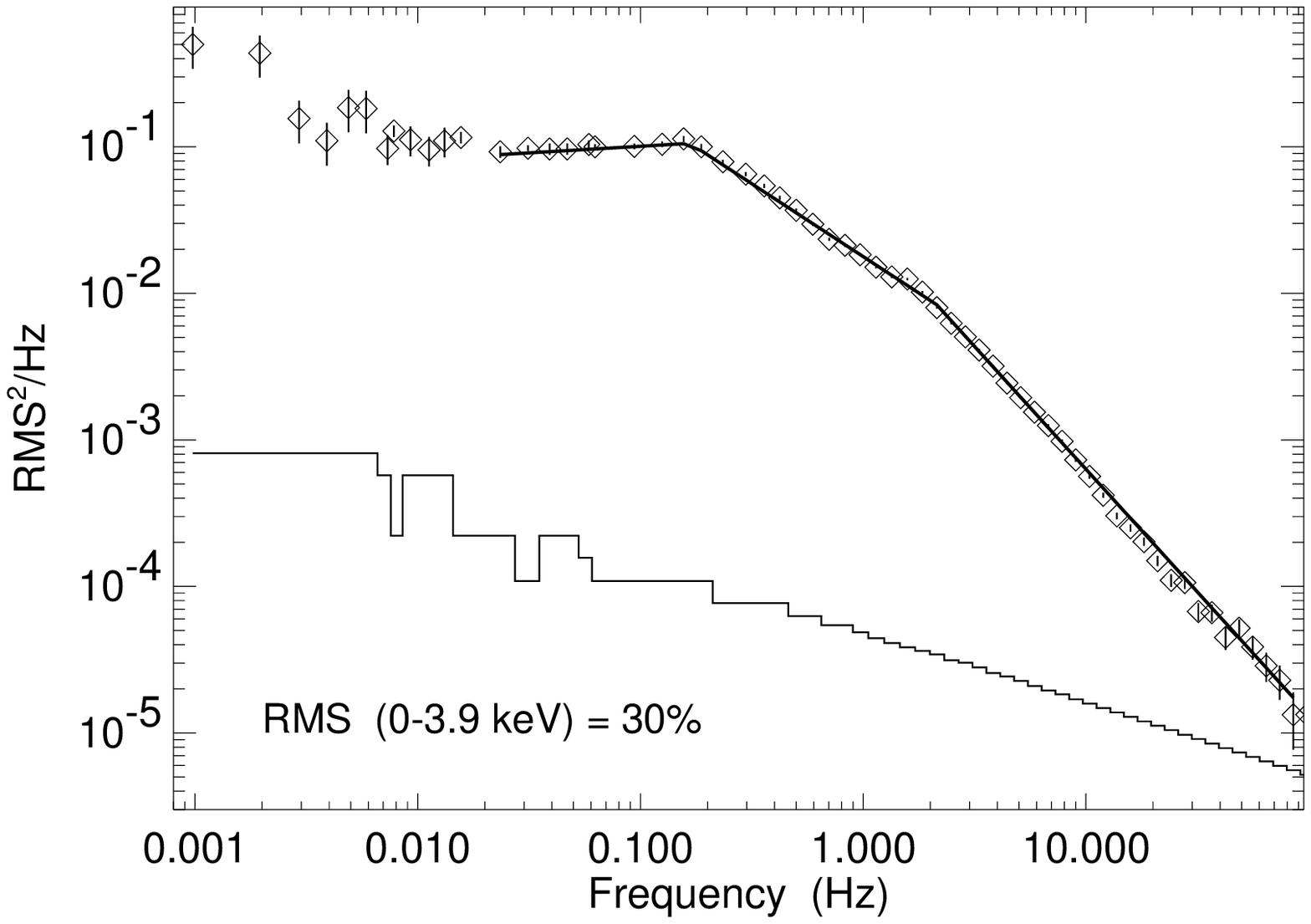,width=0.48\textwidth}
\psfig{figure=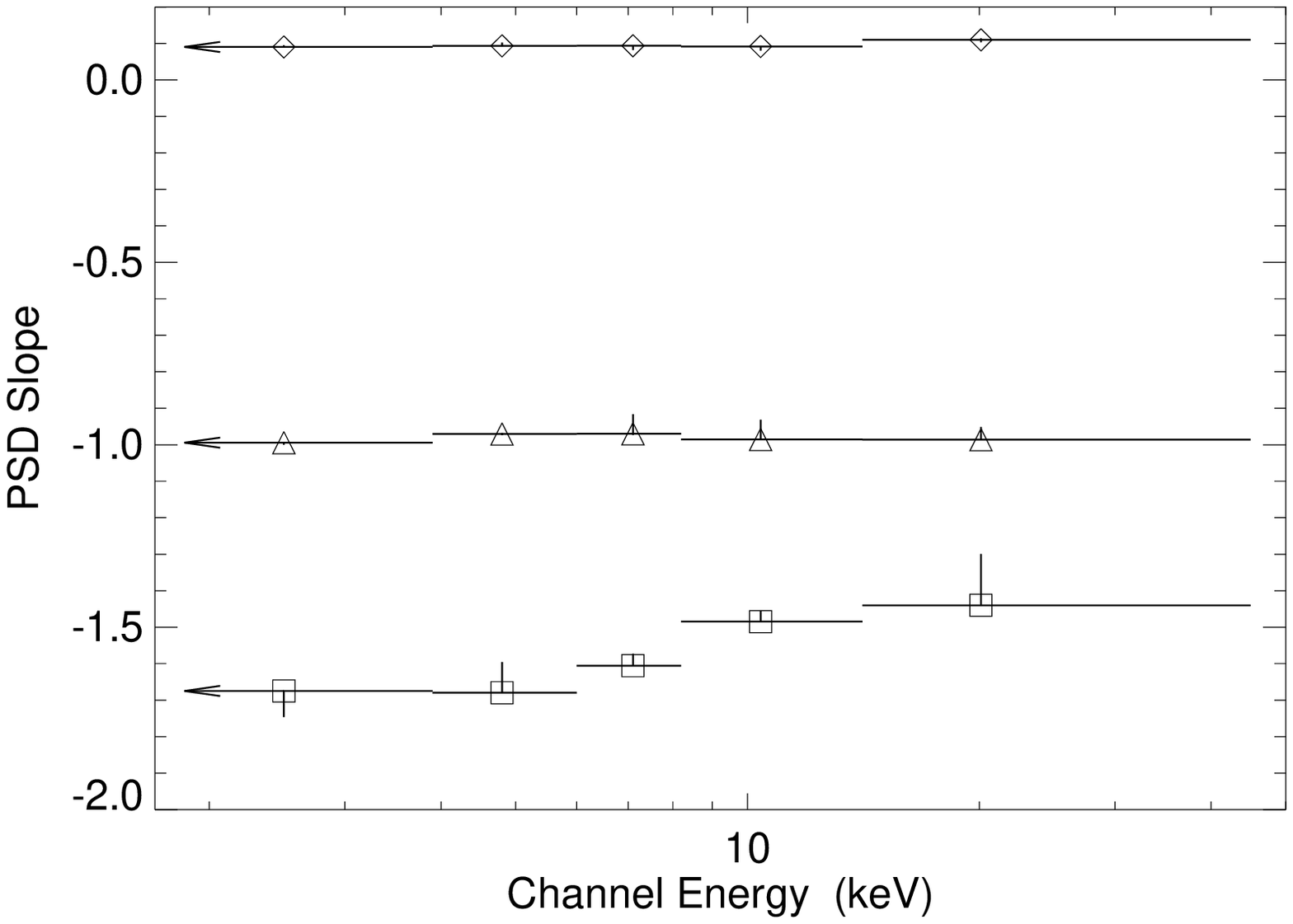,width=0.48\textwidth}
}
\caption
{{\it Left:} The same ($0-3.9$\,keV) PSD presented in Fig.~\ref{fig:psds},
except that here we also show the best fit doubly broken power law (fit in
the range $0.02-90$ Hz, reduced $\chi^2=4.0$).  {\it Right:} Best fit power
law slopes, from doubly broken power law fits, as a function of channel
energy.  Diamonds are the slopes in the $\approx 0.02-0.2$\,Hz range,
triangles are the slopes in the $\approx 0.2-2$\,Hz range, and squares are
the slopes in the $\approx 2-90$\,Hz range.}
\label{fig:bknpow}
\end{figure}

As a phenomenological description of the PSDs, we have fit doubly broken
power laws over the frequency range $0.02$--$90$\,Hz.  We show an example of
one of these fits in Figure~\ref{fig:bknpow}.  The reduced $\chi^2$ for these
fits ranged from $4$ for the lowest energy channel to $16$ for the highest
energy channel. The somewhat large reduced $\chi^2$ values result from the
excellent statistics of the data, and indicate that doubly broken power
laws are only qualitative fits to the PSDs.  Although the PSD definitely
contain structure not fit by the broken power law, Figure~\ref{fig:bknpow}
shows that they capture the essential features of the $0.02$--$90$\,Hz data.
The overall normalization of the fits decreases slightly with increasing
energy.  The low-frequency power-law index, the middle-frequency power-law
index, and the two energy breaks do not show any significant dependence
upon channel energy.  The high frequency power law index, on the other
hand, clearly `hardens' with increasing energy.  This hardening of the PSD
with increasing energy is evident in Figure~\ref{fig:bknpow}b.  We present
all of our best fit parameter values in Table~\ref{tab:bknpow}.

The results of this doubly broken power law fit are comparable to the
results found by \citey{belloni:90a} using multiple EXOSAT
observations. These authors also found the PSD to be flat below a break at
$0.04$--$0.4$\,Hz, then approximately $\propto f^{-1}$ up to a break at
$1$--$6$\,Hz, and approximately $\propto f^{-2}$ at higher frequencies.
There was weak evidence for the high frequency power law becoming less
steep with increasing energy as seen here.  Furthermore, the overall rms
variability ranged from $28\%$--$38\%$, consistent with the values
discussed here.  Our results and those of \citey{belloni:90a} are also
consistent with the trends observed by {\it Ginga}, as shown in 
\citey{miyamoto:92a}.  \citey{belloni:90a}, however, did not consider the
coherence or time delays between various energy bands.  As we shall
discuss below, these additional statistics offer the promise of
constraining physical models of this system.

\vfill\eject

\end{multicols}

\begin{table}
\caption{Parameter values for a doubly broken power law fit to the
$0.02-90$\,Hz PSD data.  The lowest frequency PSD is fit to a function of
the form $A_n f^{\alpha_1}$.  The location of the first break is at
$\beta_1$ and $\alpha_2$ is the best fit index of the second power law.
The location of the second break is at $\beta_2$ and $\alpha_3$ is the best
fit index of the third power law.  The reduced $\chi^2_r$ values for all
the fits are for $44$ degrees of freedom.  Uncertainties given are at the
90\% level for one interesting parameter ($\Delta\chi^2 = 2.7$).  Parameters
without errors have uncertainties smaller than the last shown decimal
place.}
\bigskip
\hfill\hbox{
\begin{tabular}{lllllllr}
\hline
\hline
Channel Energy & $A_n$ & $\alpha_1$ & $\beta_1$ [Hz] & $\alpha_2$ & $\beta_2$ [Hz] & $\alpha_3$ & $\chi^2_r$ \\
\hline
($0-3.9$\,keV) &$0.124\pm0.001$ & $0.090^{+0.005}_{-0.000}$ & $0.17$ & $-0.994^{+0.001}_{-0.006}$ & 2.1 & $-1.67^{+0.00}_{-0.07}$ & 4.0 \\
($3.9-6.0$\,keV) & $0.116\pm0.001$ & $0.094^{+0.008}_{-0.002}$ & $0.17$ & $-0.970\pm0.003$ & $2.1^{+0.5}_{-0.0}$ & $-1.68^{+0.08}_{-0.0}$ & 6.1 \\  
($6.0-8.2$\,keV) & $0.130^{+0.005}_{-0.027}$ & $0.092^{+0.000}_{-0.012}$ & $0.17\pm0.01$ & $-0.970^{+0.053}_{-0.003}$ & $2.2^{+0.6}_{-0.1}$ & $-1.61^{+0.04}_{-0.00}$ & 4.4 \\
($8.2-14.1$\,keV) & $0.093^{+0.003}_{-0.000}$ & $0.092^{+0.000}_{-0.011}$ & $0.19\pm0.01$ & $-0.985^{+0.054}_{-0.000}$ & $2.1^{+0.5}_{-0.1}$ & $-1.48^{+0.03}_{-0.00}$ & 9.1 \\
($14.1-45$\,keV) & $0.084^{+0.002}_{-0.000}$ & $0.110^{+0.003}_{-0.007}$ & 0.18 & $-0.986^{+0.034}_{-0.000}$ & $1.9^{+1.1}_{-0.1}$ & $-1.44^{+0.14}_{-0.00}$ & 16.2 \\
\hline
\end{tabular} }\hfill
\vspace*{3mm}
 
\label{tab:bknpow} 
\end{table}

\section*{}
\vskip -15pt

\vfill\eject

\section{Coherence Function}\label{sec:coher}

\subsection{The Coherence Function}\label{sec:simpcoh}

Coherence is a lesser known and lesser used statistic than the power
spectrum.  Like the time delay, it is defined for two or more concurrent
processes (cf. \cite{bendat,vaughan:97a}).  It is closely related to the
cross correlation function, and is particularly useful for studying
processes whose power density spectra contain features, such as QPO or
sudden changes in slope, because it provides a means of isolating for study
a particular Fourier frequency or range of frequencies that are strongly
correlated across energy bands.

The coherence function, $\gamma^2(f)$, is a Fourier-frequency-dependent
measure of the degree of \emph{linear} correlation between two concurrent
time series.  Specifically, it gives the fraction of the mean-squared
variability at $f$ of one time series that can be attributed to, or
equivalently predicted from, the other.

Consider two statistically independent time series, $s(t)$ and $h(t)$,
measured concurrently.  In this work, we are interested in cases where $s$
and $h$ are X-ray light curves measured simultaneously in two energy
channels.  It is (almost) always possible to relate $s$ and $h$ by a linear
transformation of the form
\begin{equation}
h(t) = \int_{-\infty}^\infty t_r(t-\tau)s(\tau)d\tau.
\label{tr(tau)}
\end{equation}
The function $t_r(\tau)$ is called the transfer function between
$s$ and $h$.  In the frequency domain, eq.~(\ref{tr(tau)}) becomes
\begin{equation}
H(f)=T_r(f)S(f)
\label{Tr(f)}
\end{equation}
where $S(f)$, $H(f)$, and $T_r(f)$ are the Fourier Transforms of $s(t)$,
$h(t)$, and $t_r(t)$ respectively.  Imagine making a sequence of
measurements of $s(t)$ and $h(t)$, each of finite duration.  If $T_r(f)$ is
the same for each measurement of $s$ and $h$, the processes are said to be
perfectly coherent.  That is, a measurement of one, say $s(t)$, enables us
to predict the other, $h(t)$.  The coherence function measures the degree
to which $T_r(f)$ is constant for the data segments and Fourier frequencies
over which we average.  This is equivalent to a measure of the degree of
linear correlation, in the fractional mean-squared sense, between $s$ and
$h$ at $f$.

The coherence function is defined by
\begin{equation}
\gamma^2(f) = {{\left|\left < S^\ast(f)H(f)\right>\right|^2} \over {
    \langle \left|S(f)\right|^2 \rangle \langle \left|H(f)\right|^2
    \rangle}},
\label{gamma}
\end{equation}
where triangular brackets denote an average over an ensemble of
measurements.  The one sigma uncertainty in $\gamma^2(f)$, in the case of
Gaussian statistics, is
\begin{equation}
\delta\gamma^2(f)= {{\sqrt{2}\gamma^2(f)\left[1-\gamma^2(f)\right]} \over
                    {\left|\gamma(f)\right|\sqrt{n}}},
\label{deltagamma}
\end{equation}
where $n$ is the number of measurements averaged in computing $\gamma^2(f)$
(\cite{bendat,vaughan:97a}).

The quantity within triangular brackets in the numerator, $S^\ast(f)H(f)$,
is the cross power spectrum (sometimes called the complex cross spectrum or
cross spectral density) between $s$ and $h$.  As discussed above, it is a
complex quantity whose phase is a measure of the shift between the two
light curves at $f$.  The denominator in eq.~(\ref{gamma}) is the product
of the average power spectra.

An intuitive feel for the coherence function may be gained by looking at it
geometrically.  Let the $j$th component of the FFT of $s$ (where we now
consider an equispaced, discrete-time process $s_k$) be $S_j$, which we
write as $S_j=|S_j|e^{i\phi_s}$.  The product $S^\ast_jH_j$ is then equal
to $|S_j||H_j|e^{i(\phi_h-\phi_s)}$.  The coherence function is defined as
the squared magnitude of the average value of this product,
$\left|\left<S^\ast_jH_j\right>\right|$, normalized by the the product of
the individual power spectral densities, $\langle \left|S_j\right|^2
\rangle \langle \left|H_j\right|^2 \rangle$.

\bigskip
\centerline{
\epsfxsize=0.90\hsize {\epsfbox{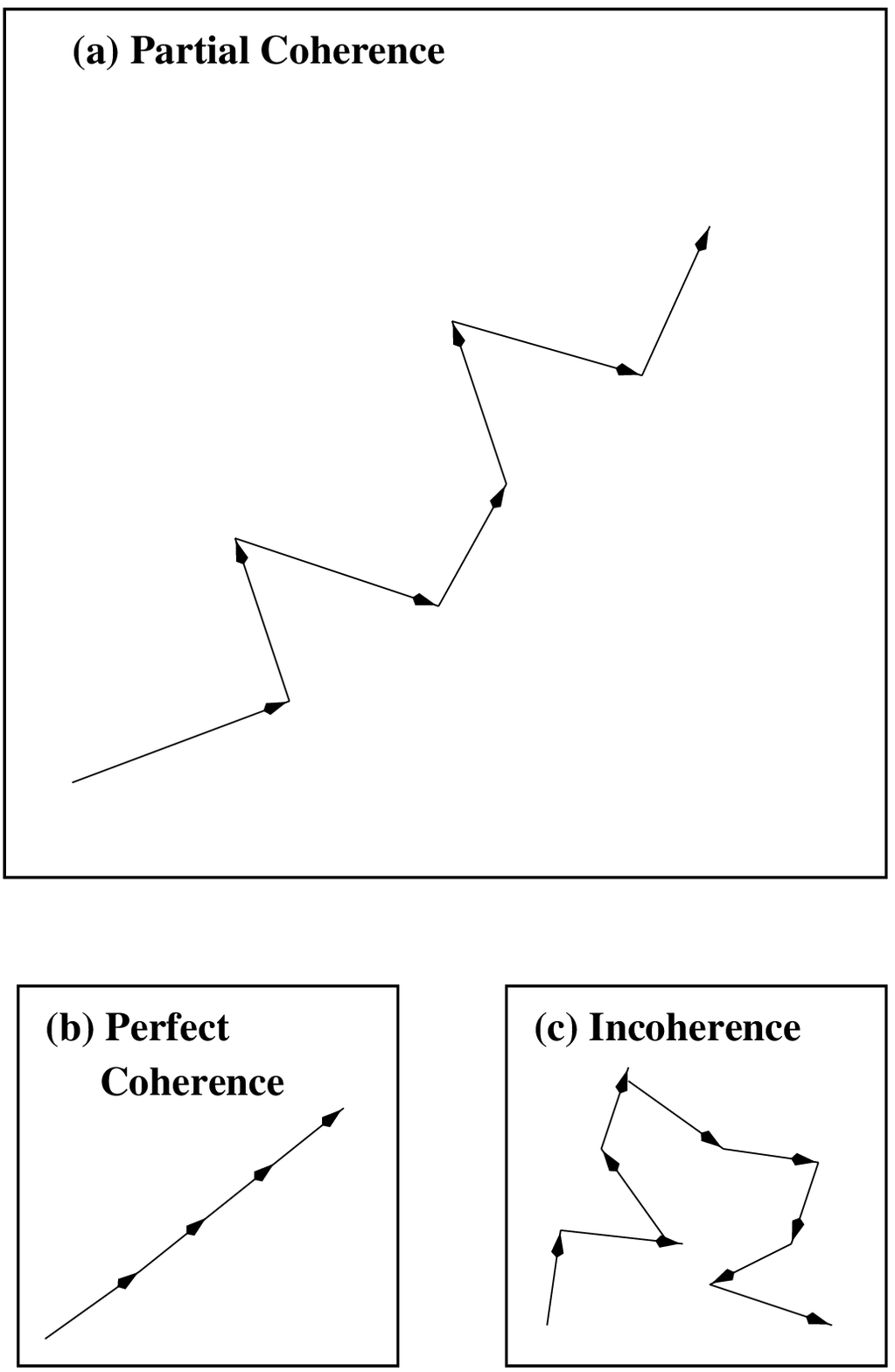}}
}
\medskip
\figcaption {a) Vector illustration of the coherence function.  Each vector
is the complex number $A^\ast B$ at a single Fourier frequency $f$ for a
single measurement of $A$ and $B$.  If the two processes are coherent, then
$A^\ast B$ will always have the same phase, as shown in (b).  The vectors
will be perfectly aligned and $|\sum A^\ast B|$ will be equal to
$\sum|A^\ast B|$.  If the processes are incoherent, then the phase of
$A^\ast B$ will be uniformly and randomly distributed on $[-\pi,\pi]$, as
illustrated in (c).  In this case, $\langle |A^\ast B|^2 \rangle /( \langle
|A|^2 \rangle \langle |B|^2 \rangle )\rightarrow 0$ as the number of
measurements approaches infinity.}
\label{fig:vector}
\bigskip

The process of finding the average value $S^\ast_jH_j$ may be pictured as a
vector sum in the complex plane.  The sum may be found geometrically by
drawing each term in the sum as a vector and adding them head to tail, as
in usual vector addition, as shown in 
Figure~3a.  
If the time series are coherent at frequency $f_j$, then the phase
difference between them, and hence the direction of the vector product
$S^\ast_jH_j$, will be the same for each measurement in the sequence, as
shown in Figure~3b.  
Geometrically, the vectors will all point in the same direction, and the
sum of the vectors will have the same magnitude as the scalar sum of their
magnitudes.  In this case, the numerator in eq.~(\ref{gamma}) will equal
the denominator, and the coherence is unity.  Coherence can only be
computed over an ensemble of measurements.

In the opposite extreme, the soft and hard channels are completely
incoherent.  The phases of $S_j$ and $H_j$ are unrelated, and
$\phi_h-\phi_s$ will be uniformly and randomly distributed on $[-\pi,\pi]$.
This is just a 2-D random walk, as illustrated in Figure~3c.
As the number of observations increases, the measured coherence goes to
zero.

The light curves $s(t)$ and $h(t)$ need not look at all alike to be
perfectly coherent.  Indeed, they will look alike only if the time
difference at each Fourier frequency, $\delta t=(\phi_h-\phi_s)/(2\pi f)$,
is constant, which is equivalent to $t_r(\tau)=k\delta(\tau-\delta t)$,
where $k$ is a scaling factor between $s$ and $h$.

An example of near perfect coherence can be found in the data studied for
this paper.  Figure~4 
shows lightcurves of \cyg in the ($0$--$3.9$\,keV) and ($14.1$--$45$\,keV)
energy bands.  We have divided the data into 128 s segments for measuring
coherence.  The average direction in the complex plane of the Fourier
transform of each segment over the Fourier frequency range $1.5$--$2.5$\,Hz
is shown above the top light curve and below the bottom, labeled
$\left<A\right>$ and $\left<B\right>$.  The product $\left <A^\ast B
\right>$ is shown between the light curves.  Notice that the phase of the
Fourier transform varies randomly from segment to segment.  In contrast,
$\left < A^\ast B \right >$ remains constant, indicating a fixed phase
shift between the light curves in the 1.5--2.5 Hz range.  As we will show
later, the coherence is unity over a broad frequency range, even though the
phase shift varies by a factor of 4 over the same range.

In practice, measurement noise (in our case Poisson counting noise) causes
the measured coherence to be always smaller than unity.  Poisson noise
contributes a random part to each phase measurement, causing the $\left <
S(f)^\ast H(f) \right >$ to be slightly misaligned.  For weak signals
(relative to the noise), counting noise dominates the signal.  Poisson
noise always dominates at sufficiently high frequencies because it has a
(nearly) white power density spectrum.  Coherence is also difficult to
measure at low frequencies because the number of independent measurement is
smaller.  Vaughan \& Nowak (1997) discuss methods to correct $\gamma^2$ and
$\delta\gamma^2$ for counting noise.

\subsection{Coherence in \cyg}

\begin{figure}
\centerline{
\psfig{figure=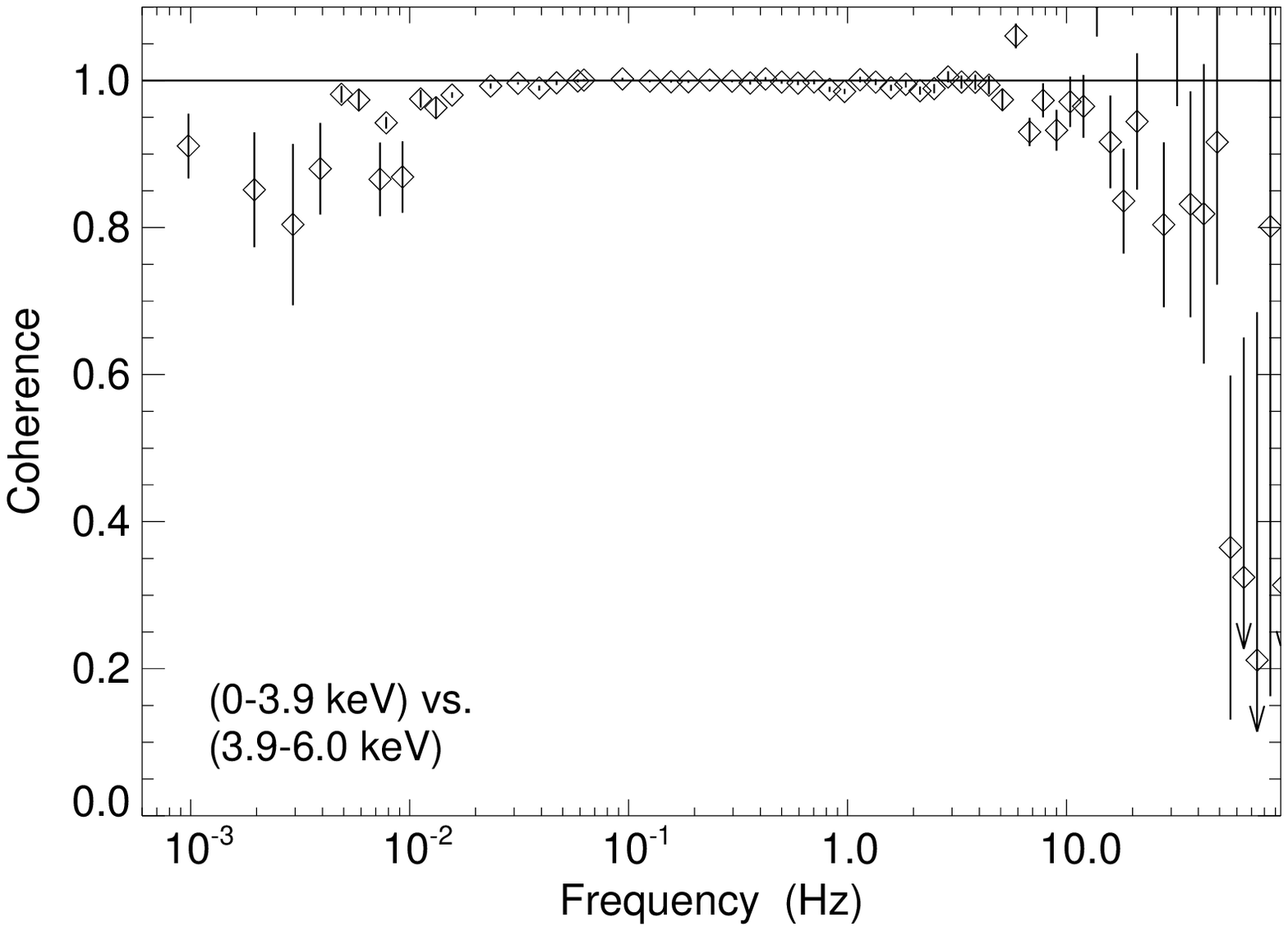,width=0.456\textwidth}
\psfig{figure=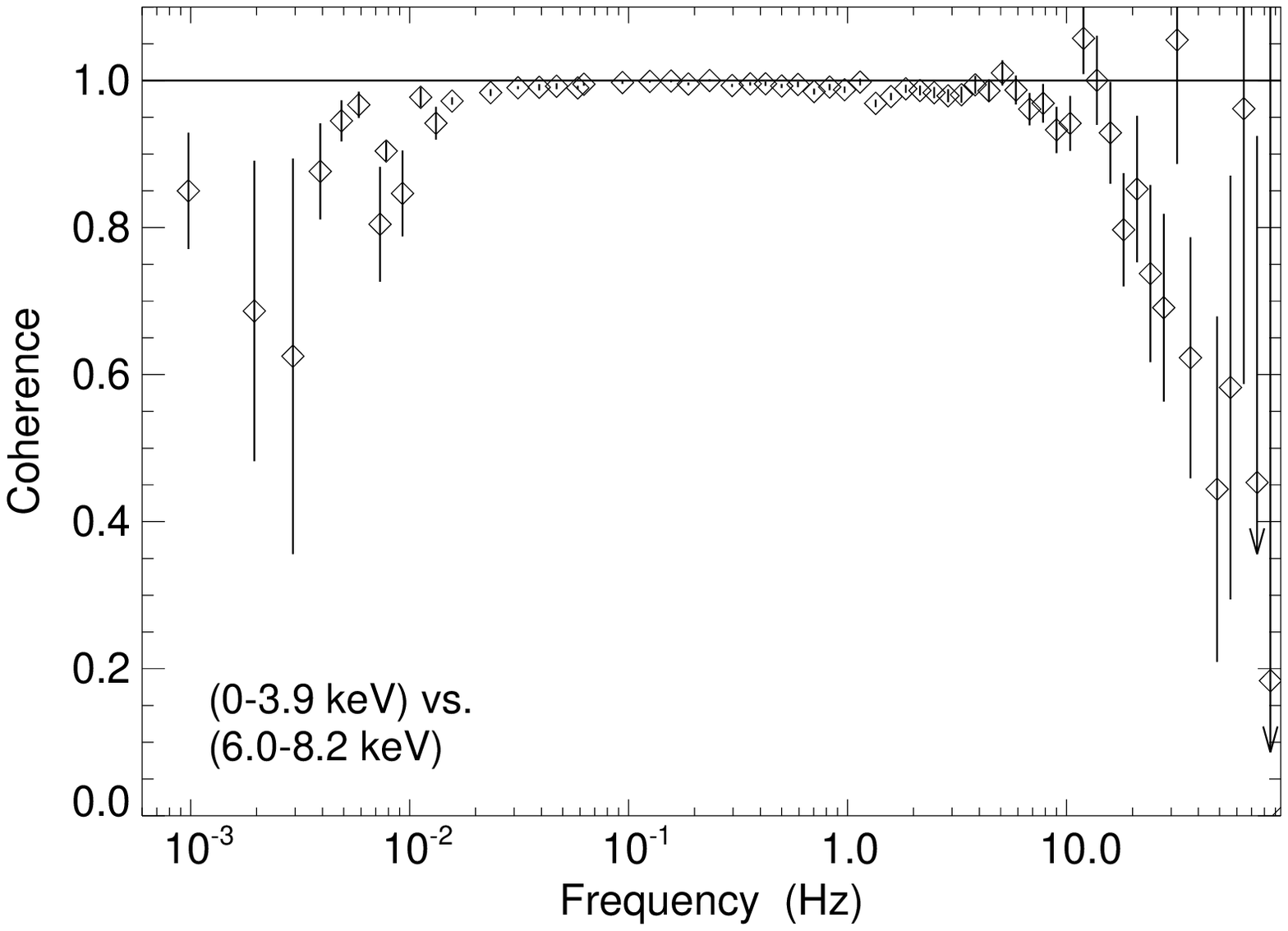,width=0.456\textwidth}
}
\centerline{
\psfig{figure=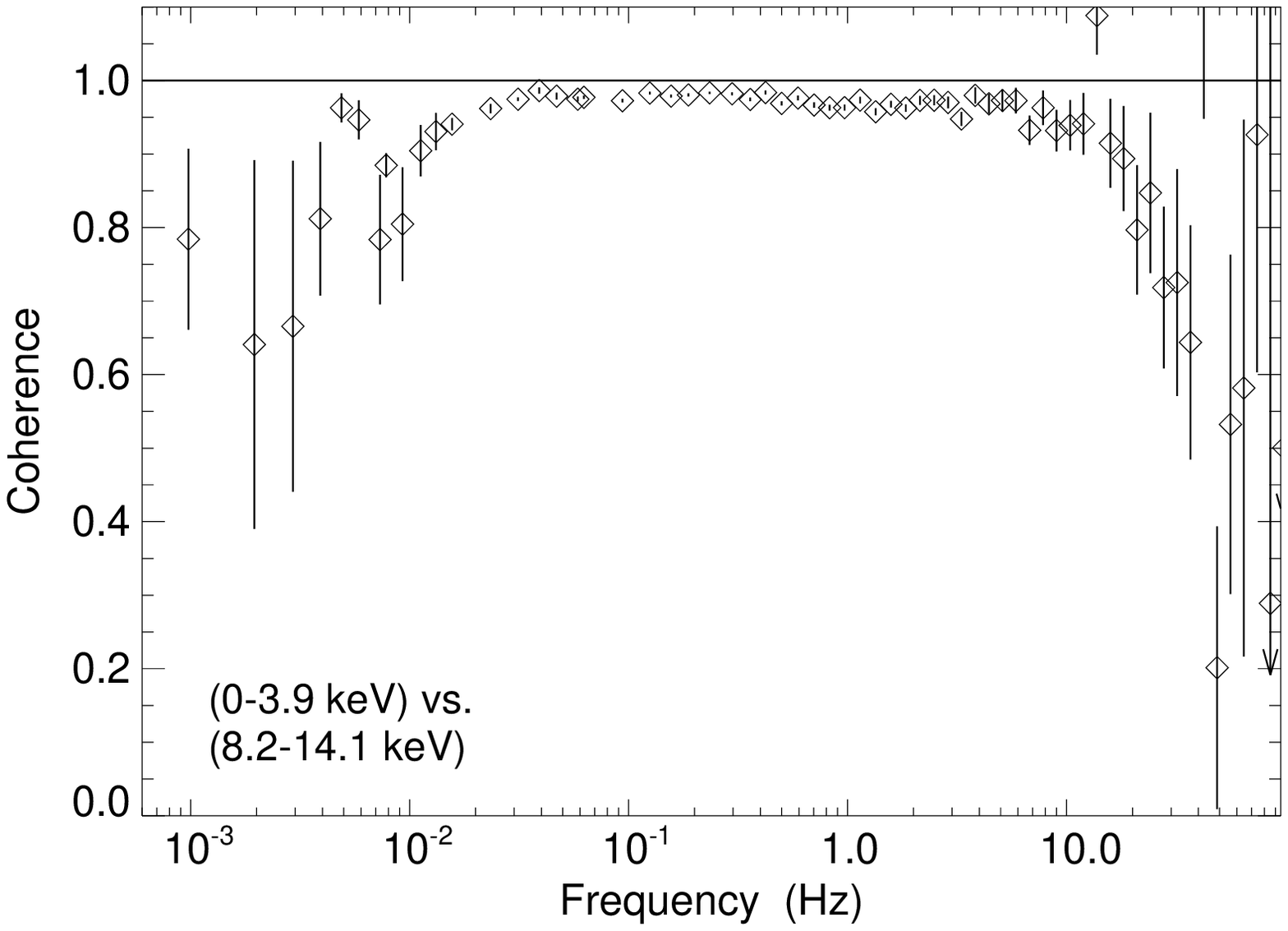,width=0.456\textwidth}
\psfig{figure=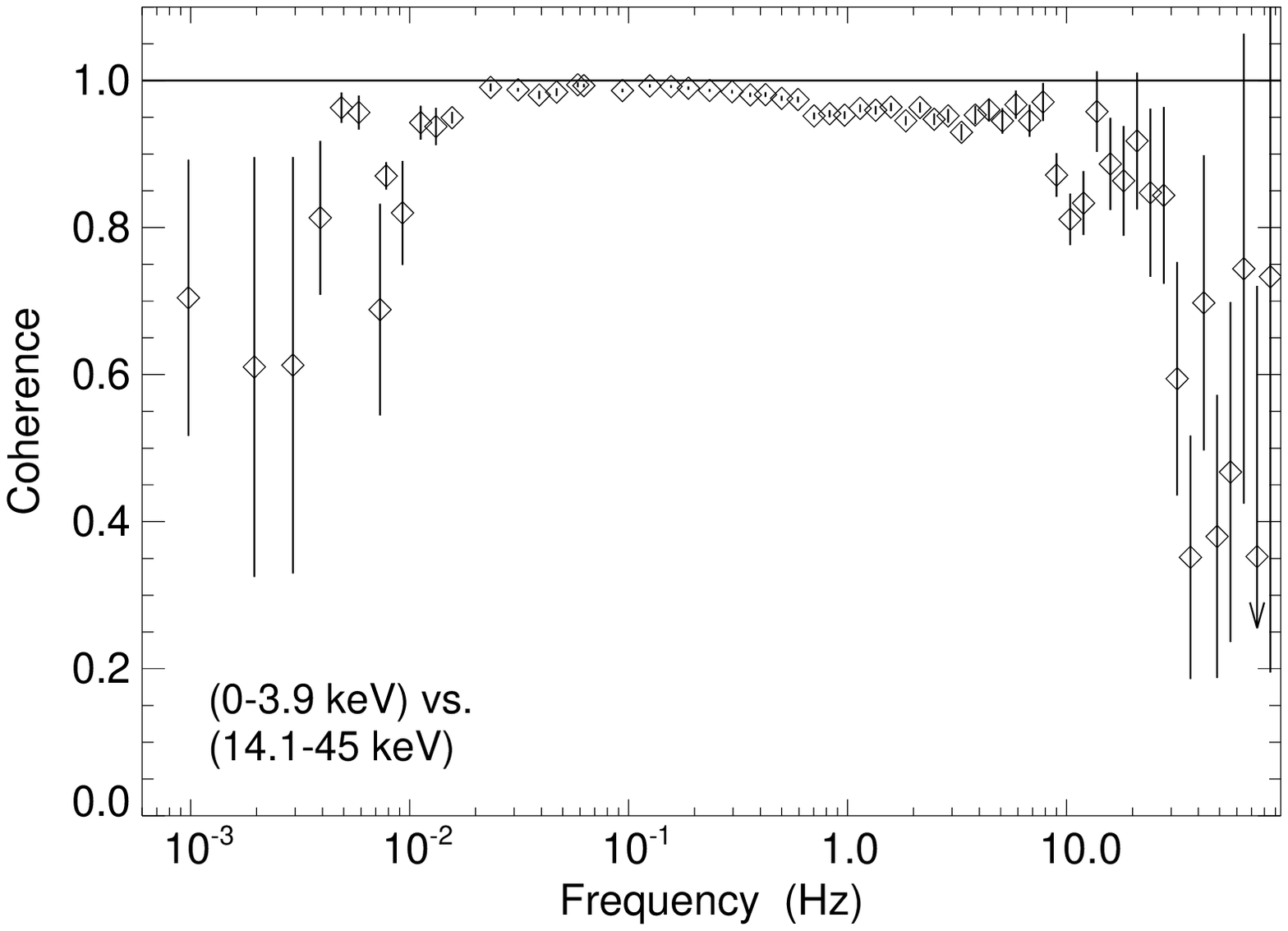,width=0.456\textwidth}
}
\caption
{Coherence, as a function of Fourier frequency, for various energy bands
vs. the lowest energy band (0$-$3.9\,keV).  (Solid line is coherence
equals unity.) Error bars are calculated from eq.~(\ref{deltagamma}).}
\label{fig:cohdata}
\end{figure}

In Figure~\ref{fig:cohdata} we present the coherence function for our \cyg
observation.  Specifically, the coherence function and its error bars were
calculated by applying eq.~(8) of \citey{vaughan:97a} to our data.  (Strictly
speaking that equation is most valid for the high coherence, high signal
regime of our PSDs, which here is $f \aproxlt 10$\,Hz .  Improved estimates
using the more complicated formulae presented in \citey{vaughan:97a} are
unlikely, however, to drastically change our results.)  All comparisons
shown are to the ($0-3.9$\,keV) energy band. As discussed by
\citey{vaughan:97a}, the expectation from many theoretical models is that
the coherence should be less than unity.  As with previous observations of
\cyg (\cite{vaughan:91a,vaughan:97a,cui:97b}), the coherence is remarkably
close to unity over a wide frequency range, even for rather disparate
energy bands.  Over the range $0.02-10$\,Hz, the coherence is nearly unity.
In the two highest energy channels, there is a trend for the coherence to
drop slightly above $\approx 1$\,Hz.  In all energy channels, there is a
trend for the coherence to drop below $\approx 0.02$\,Hz and above $\approx
10$\,Hz.  For the most part, the deviation from unity coherence becomes
greater with increasing energy.  At the lower frequency end, the coherence
recovers back towards unity near $0.005$\,Hz, coincident with the weak low
frequency feature seen in the PSD.

Further of note is that in the $0.02-0.2$\,Hz range, 
energy band 5 is actually slightly more coherent than 
energy band 4 (both as compared to energy band 1).  Due to
the extremely good measurements in this frequency range, this difference is
actually significant.  The frequency regime $0.02-0.2$\,Hz corresponds to
the flat part of the PSDs for energy bands $1$--$5$.

We highlight these features in Figure~6 
by presenting a grey scale plot of
$\log[1-\gamma^2(f)]$, which enhances the small differences from unity
coherence.  In this figure, high coherence is white, and low coherence is
black. The lowest energy channel, by definition, is coherent with itself.
The general trend to lose coherence at both low and high frequencies is
evident in this plot, as is the trend for this loss to increase with
energy.  The highest energy channel does indeed appear more coherent in the
$0.02-0.2$\,Hz range.  Note that the high coherence coincident with the
$0.005$\,Hz feature is also evident in Figure~6.

\smallskip
\centerline{
\epsfxsize=0.96\hsize {\epsfbox{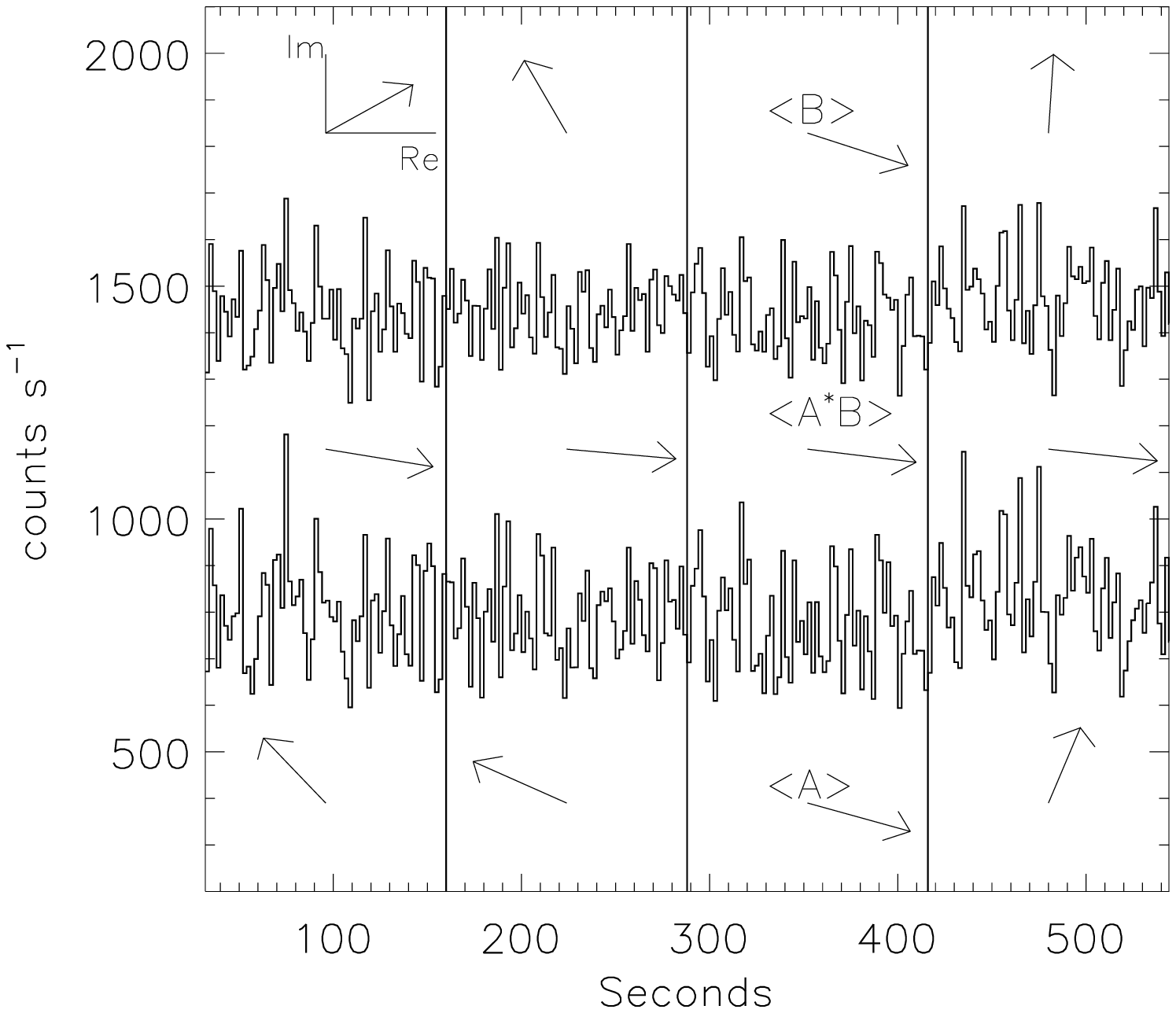}}
}
\bigskip
\figcaption {Lightcurves (solid lines) for a portion of our \cyg data.  The
top curve is the ($14.1$--$45$\,keV) energy band, shifted by 600 ${\rm cts~
s}^{-1}$, while the bottom is for the ($0$--$3.9$\,keV) energy band.  Solid
vertical lines divide the data into segments of 128\,s.  Arrows represent
the {\it directions} in the complex plane of the FFT and cross power
spectral density for each data segment, averaged over the frequency range
$1.5$--$2.5$\,Hz. Poisson noise has {\it not} been subtracted; therefore,
it is expected that $\langle A^\ast \rangle \langle B \rangle \ne \langle
A^\ast B \rangle$, as seen for these data.}
\label{fig:funky}
\bigskip

\end{multicols}
\section*{}
\vskip -15pt

\vfill\eject

\centerline{
\epsfxsize=0.96\hsize {\epsfbox{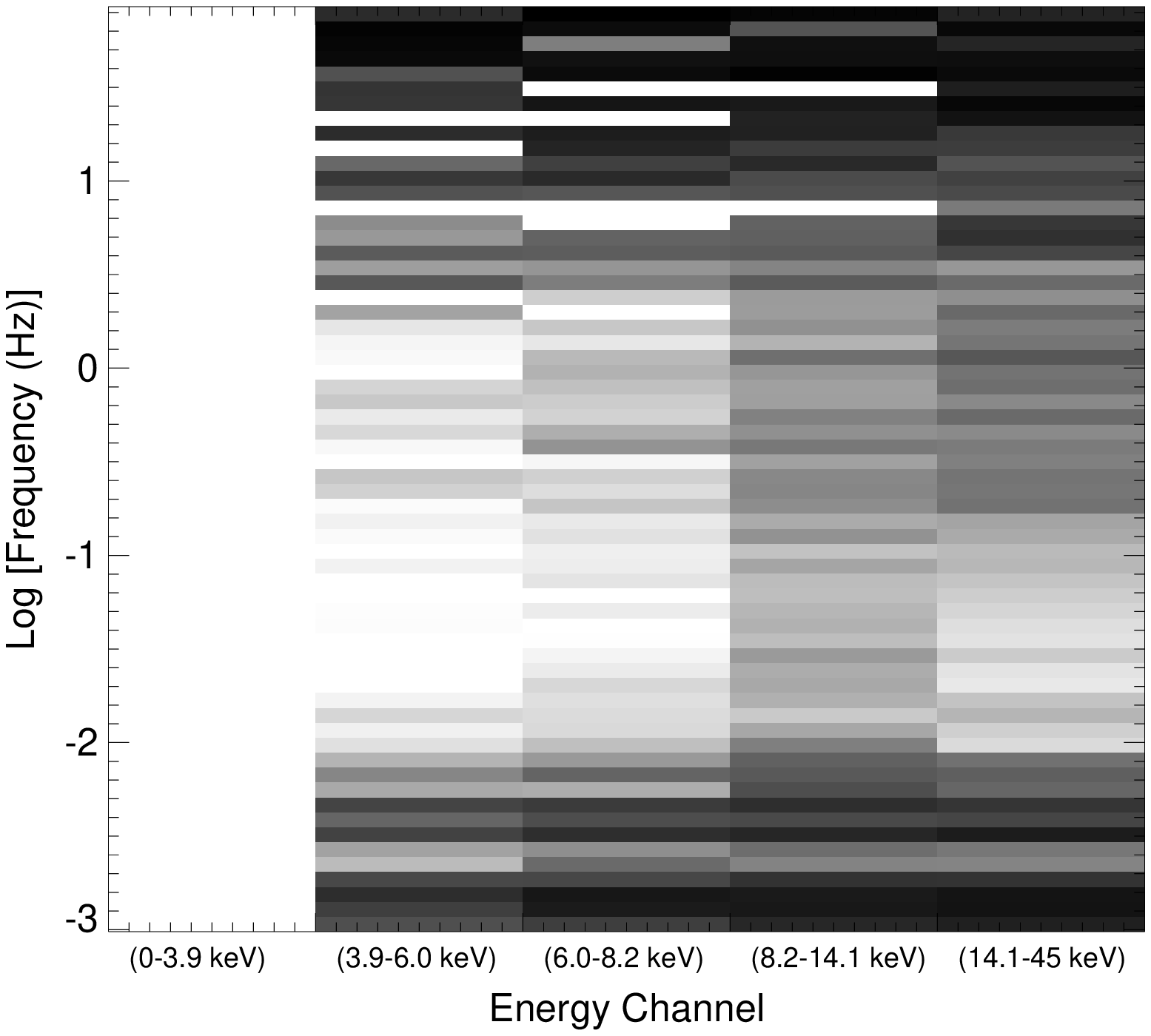}}
}
\bigskip
\figcaption {Grey scale representation of the coherence function as a
function of frequency (vertical axis) and energy band (horizontal axis).
Specifically, there are 256 grey scale levels corresponding to the value of
$\log[1 - \gamma^2(f)]$, where $\gamma^2(f)$ is the noise-subtracted
coherence function for the indicated energy band as compared to the
($0-3.9$\,keV) energy band.  Values of $\gamma^2(f)>0.999$ were set to $0.999$.
White corresponds to the highest coherence levels, black to the lowest
coherence levels.}
\label{fig:tv}
\bigskip

The deviations to greater than unity for the noise-subtracted coherence in
Figures~\ref{fig:cohdata} and 6
are due to both statistical {\it and} systematic errors.  As the coherence
function is a fourth order statistic [as compared to a second order
statistic like the PSD, cf. \citey{bendat}], it is very sensitive to
uncertainties in the noise subtraction.  With a systematic uncertainty in
the deadtime as large $\aproxlt 25\%$, and even with the background
variability levels as seen in Figure~\ref{fig:psds}, we are still \emph{very}
confident of the coherence estimates from $\approx 10^{-3}$--$30$\,Hz,
where systematic uncertainties are essentially negligible.  Between
$\approx 30$--$100$\,Hz, our estimates are that systematic uncertainties
increase the shown error bars by no more than $\approx 30\%$.  The
noise-subtracted coherence points do show that our noise subtraction is not
perfect; however, the general trend of coherence loss is likely fairly
secure to $\approx 100$\,Hz.

\section{Time Lags}\label{sec:tlinterp}

\subsection{Simple Considerations}\label{sec:tlsimp}

The Fourier time delay, or lag, is also computed for two concurrent and
correlated time series.  Like the coherence, it is related to the cross
correlation, and is a Fourier-frequency dependent measure of the time delay
between the time series (cf. \cite{miyamoto:89a,miyamoto:92a,vanderklis:89b}).

As in \S\ref{sec:coher}, let $s(t)$ be a discretely and uniformly sampled
``soft energy'' lightcurve and $h(t)$ be a discretely and uniformly sampled
``hard energy'' lightcurve.  These lightcurves, as for the PSD
(cf.\ \S\ref{sec:psd}) and coherence (\S\ref{sec:coher}), are divided into
data segments of uniform length and an FFT is performed on each segment.  The
Fourier phase lag, $\phi(f)$, is the phase of the {\it average} cross
power spectrum.  That is, $\phi(f)={\rm arg}[C(f)]$, where
\begin{equation}
C(f)=\langle S^*(f) H(f) \rangle ~~.
\label{eq:lag}
\end{equation}
The Fourier time lag is constructed from $\phi(f)$ by dividing through by
$2\pi f$, i.e. $\tau(f) \equiv \phi(f)/2\pi f$.  The Fourier time lag can
be either positive or negative.  For our sign convention, a positive time
lag indicates that the hard light curve lags the soft light curve.  Note
that the Fourier phase is defined on the interval $(-\pi,\pi)$, hence a lag
of $\pi/2$ in reality could be a lead of $3\pi/2$.  For a further
discussion, see \citey{nowak:96a}.

In general, both the phase lag and time lag are non-constant functions of
Fourier frequency, $f$.  To get a feel for what this can mean, consider the
following simple example.  Imagine we have a source of fluctuations that
produces soft X-ray photons some distance from a region that also responds
to these fluctuations by producing hard X-ray photons.  If the fluctuations
can propagate from the soft-photon-producing region to the
hard-photon-producing region, without dispersion, then we expect a time
delay between the soft and hard photons that is {\it independent of Fourier
frequency}, $f$.  The time delay at all Fourier frequencies will simply be
the distance between the soft X-ray source and the hard X-ray response
divided by the propagation speed. This means that the Fourier phase lag,
$\phi(f)$, will increase linearly with $f$ (modulo integer multiples of
$2\pi$).

If the two processes are perfectly correlated (cf. \S\ref{sec:coher}), the
two processes are related by a constant, linear transfer function.  The
time lags observed in \cyg are not independent of Fourier frequency;
however, the phase lags are roughly independent of $f$.  Let us look at a
crude but illustrative transfer function that produces a constant phase
lag.  Specifically, we consider a transfer function, $T_r(f)$, with both
constant phase and amplitude.  This simple example is in fact a reasonable
approximation to {Cygnus~X--1}.  In this case, we can write
\begin{equation}
H(f) ~=~ A \exp(\pm i \Delta \phi) ~ S(f) ~ ~, 
\end{equation}
where both $A$ and $\Delta \phi$ are real constants.  The plus sign is for
$f > 0$ and the minus sign is for $f < 0$, because the
Fourier transform of a real time series satisfies $H(f) = H^*(-f)$. Taking the
inverse transform of $A \exp(\pm i \Delta \phi)$, the transfer function,
$t_r(\tau)$, can be written in the time domain as 
\begin{equation}
t_r(\tau) ~=~ A \left
[ \cos(\Delta \phi) \delta(\tau) + {{\sin(\Delta \phi)}\over{\tau}} \right
] ~~.
\label{eq:cptrans}
\end{equation}
That is, a fraction $A \cos(\Delta \phi)$ of the hard variability is
exactly coincident with the soft variability, while a (typically smaller)
fraction is delayed from the soft variability with a $\tau^{-1}$ ``tail''
in the transfer function.  For $A\sim 1$ and $\Delta \phi \sim 0.1$
radians, most of the soft and hard lightcurves are exactly coincident with
one another, and only a small fraction of the hard photons lag behind the
soft.  To illustrate the above points, we present examples of both constant
(as a function of Fourier frequency) time lags and constant phase lags in
Figure~7.

The time delays between hard and soft photons may be due to a physical
separation between a source and a response, as in the first example, or
they may be related to the emission process itself.  For the case of
Comptonization, hard photons undergo more scatterings than soft photons and
therefore are naturally delayed from soft photons.  This latter process
occurs on very short timescales (on the order of the light crossing time of
the Comptonizing region) and represents the {\it minimum} expected time
delay between hard and soft X-rays within the context of this particular
model (\cite{miyamoto:89a,miller:95a,nowak:96a,kazanas:97a}).  On the other
hand, the {\it maximum} expected time delay between hard and soft X-rays
will be less than the size of the emitting region divided by the {\it
slowest} propagation speed in that region (whether it be a sound speed or a
thermal wave speed, etc.).  As discussed below, most of the hard and soft
photons seen in \cyg are nearly temporally coincident, and therefore the
(possibly extended) sources of soft and hard X-rays are likely to be nearly
spatially coincident as well.  The more spatially coincident the soft and
hard X-ray sources/responses are, the shorter the observed time delays will
be.

\centerline{ \epsfxsize=0.96\hsize {\epsfbox{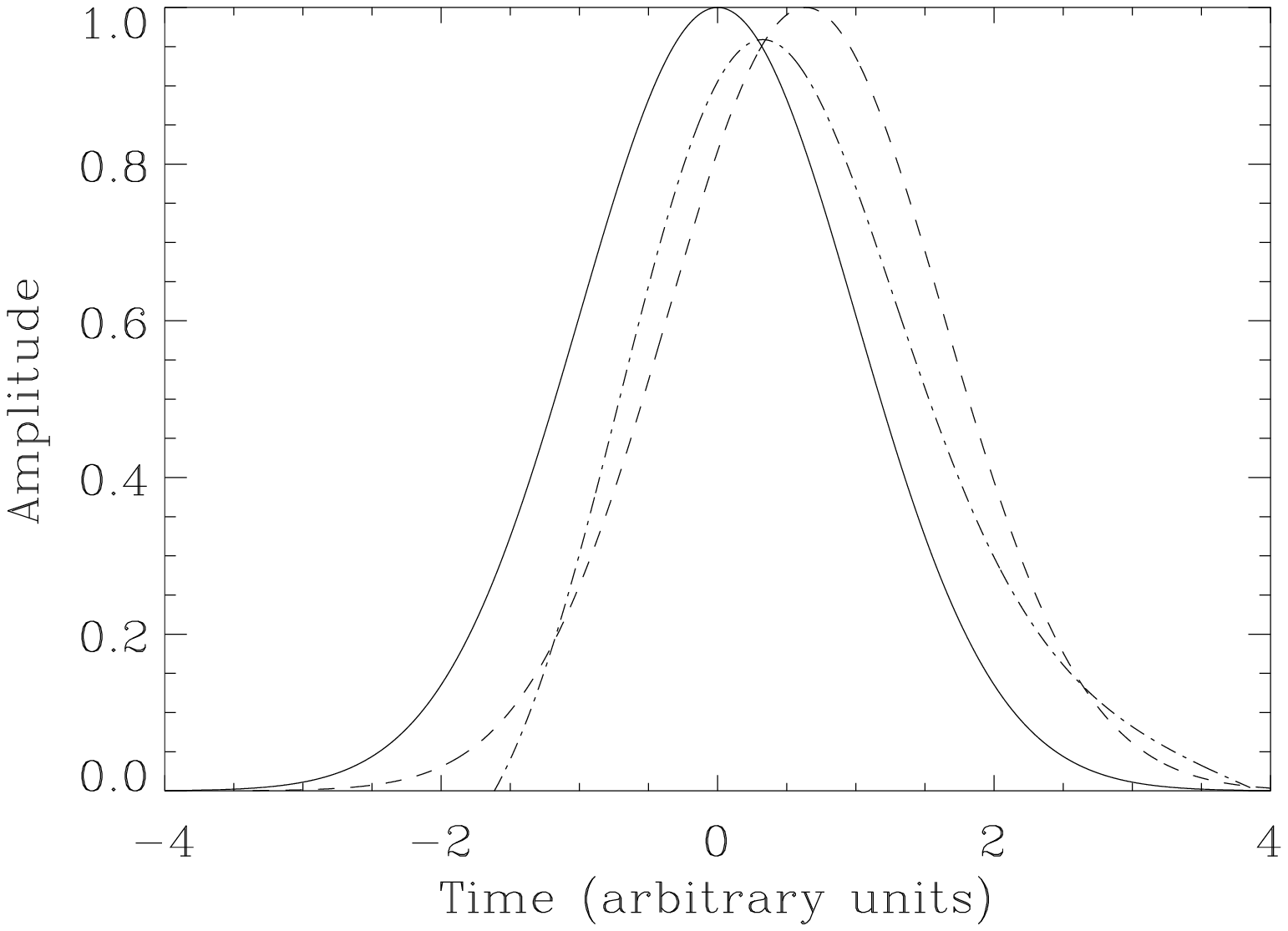}} }
\bigskip
\figcaption {Gaussian of unit amplitude and variance (solid curve), and same
Gaussian shifted in time by $\Delta t=0.5$ units (dashed line) 
and in phase by $\Delta\phi = 2\pi f_0 \Delta t$, (dot-dashed line)
where $f_0$ is the lowest Fourier frequency (here $f_0=1/8$).  Notice
that both cause a shift to later times, while the constant phase shift
also induces an extended tail.}
\label{fig:gauss}
\bigskip

\subsection{Noise Limits on Time Lag Measurements}\label{sec:lagnoise}

A geometrical interpretation of the phase/time lags is helpful in
elucidating their nature (cf. \S\ref{sec:simpcoh}).  To understand the
effects of counting noise on measurements of the time lags, we again
consider the Fourier components of a measured process as vectors in the
complex plane.  The vectors consist of the sum of a signal part and a noise
part.  Let the measured Fourier components in the soft and hard channels be
given by $S(f) \equiv A_{\rm s} (f) f^{-\alpha_s} + A_{\rm
s}^{(N)} (f)$ and $H(f) \equiv A_{\rm h} (f) f^{-\alpha_h} + A_{\rm
h}^{(N)} (f)$.  Here $A_{\rm s}$ and $A_{\rm h}$ are complex quantities
representing the signal that have a constant amplitude, but have phases
(i.e. directions in the complex plane) that are a function of frequency,
$f$. The quantities $A_{\rm s}^{(N)}$ and $A_{\rm h}^{(N)}$ are the noise
components of the Fourier transforms. They have a well defined mean
amplitude but a completely random phase.  The phases of the noise
components--- and thus their orientation in the complex plane--- vary
randomly from frequency to frequency and from data segment to data
segment. For simplicity, we shall take $A_{\rm s}(f) f^{-\alpha_s} = A_{\rm
h}(f) f^{-\alpha_h} = A(f) f^{-\alpha}$ and $|A_{\rm s}^{(N)}(f)| = |A_{\rm
h}^{(N)}(f)| = \sqrt{P_N}$, which is approximately true for our \cyg data.

As for the coherence function, when we average the CPD over discrete
samples of the lightcurve and/or over frequency bins, we are essentially
summing vectors in the complex plane.  For unity coherence, the signal
vectors each have constant directions set by the phase lags and amplitudes
set by the PSD (here equal to $A^2 f^{-2 \alpha}$).  More generally, if we
sum the signal over $N$ lightcurve samples and/or frequency bins, the total
length of the summed signal vectors becomes
\begin{equation}
N ~\sqrt{\gamma^2(f)} ~|A|^2 ~f^{-2 \alpha}  ~~,
\end{equation}
where the factor of $\sqrt{\gamma^2(f)}$ takes account of the coherence
function (see \S\ref{sec:coher}).  In our data set, however, the coherence
function is unity below $\sim 10$ Hz, and falls off rapidly at higher
frequencies.

The noise vectors associated with the cross power spectral density each
have a mean amplitude of $2 \sqrt{P_N} |A| f^{-\alpha}$ (i.e. twice the
product of the amplitude of an individual noise vector with the amplitude
of an individual signal vector).  Their phases, however, are completely
random.  Thus, in the averaging process the sum of these noise vectors
looks like a random walk in the complex plane (\cite{vaughan:91a}, and
\S\ref{sec:simpcoh} above).  The root mean square amplitude of the sum of
the noise vectors is therefore
\begin{equation}
\sqrt{2N~P_N}  ~|A| ~f^{-\alpha} ~~.
\end{equation}
The fact that this sum has a random phase leads to the uncertainty in the
phase lag determination.  In the small angle approximation (appropriate for
large $N$), the uncertainty of the phase lag is then of order the expectation
value of the magnitude of the sum of the noise vectors divided by the mean
magnitude of  the sum of the signal vectors.  We have:
\begin{equation}
\Delta \phi(f) ~\approx~ \sqrt{{P_N} \over {N~\gamma^2(f)}} ~
{{f^{\alpha}}\over{|A|}} ~~.
\end{equation}

\centerline{
\epsfxsize=0.96\hsize {\epsfbox{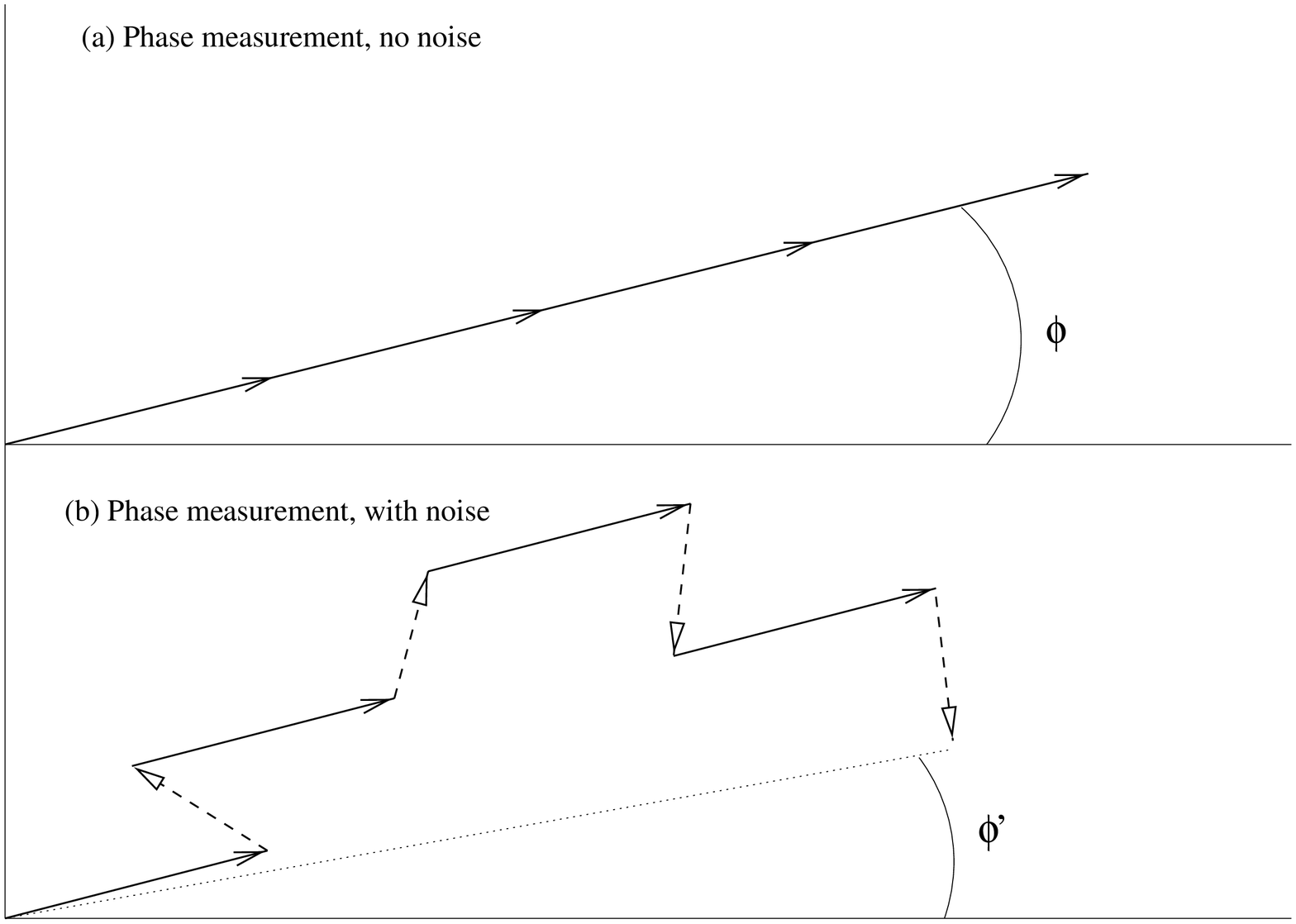}}
}
\bigskip
\figcaption {The process of measuring a phase lag in the presence and
absence of noise, illustrated as a vector sum in the complex plane for the
case of perfect coherence.  Signal vectors are displayed as solid lines in
(a) and (b), and noise vectors as dashed lines in (b).  For perfect
coherence, all signal vectors are aligned, as seen in (a) and (b).  The
noise vectors in (b) have phases uniformly and randomly distributed on
$[-\pi,\pi]$, and contribute component to the vector sum with random phase
and an amplitude that grows as the square root of the number of
measurements.  In contrast, the amplitude of the sum of signal vectors grows as
the number of measurements.}
\label{fig:lagvec}
\bigskip

The process of measuring a phase lag in the presence and absence of noise
is illustrated in Figure~8
for the case of perfect coherence.  All signal vectors, displayed with
solid lines, are aligned.  The noise vectors, displayed as dashed lines,
have phases uniformly and randomly distributed on $[-\pi,\pi]$.  The noise
vectors contribute a random component to the vector sum.  The noise
component has random phase, and an amplitude that grows as the square root
of the number of measurements.  In contrast, the amplitude of sum of signal
vectors grows as the number of measurements.

Typically, one has total observing time $T$ segmented into lightcurves of
length $T_s$ (yielding a minimum Fourier frequency $1/T_s$).  The FFTs are
also logarithmically binned over a frequency interval with {\it fractional}
width $d$. The total number of Fourier components averaged at frequency $f$
is then just the total number of lightcurves, $T/T_s$, times the number of
frequency bins averaged, $df/T_s^{-1}$, yielding $N = dTf$.  For the PSD
normalization used in \S\ref{sec:psd}, $|P_N| \approx {2/{\cal R}}$,
independent of frequency.  We then have
\begin{equation}
\Delta \phi(f) ~\approx~ 2~ \left [ d T {\cal R} \gamma^2(f) \right ]^{-1/2}
       ~ |A|^{-1} ~f^{\alpha -1/2} ~~.
\label{eq:noise}
\end{equation}
For our \cyg data $\gamma^2(f) \approx 1$ and $\alpha \sim 0.5$ (i.e. PSD
$\propto f^{-1}$) in the frequency range $\sim 0.2-2$ Hz; therefore,
$\Delta \phi(f)$ is approximately constant.  Above $\sim 2$ Hz, $\alpha
\sim 0.7-0.9$ and $\gamma^2(f)$ begins to drop rapidly.  It therefore
becomes difficult to measure phase/time lags above $\sim 30$ Hz.  This
situation is not easily improved, even with {\it substantially} longer
integration times.

As discussed below, the phase lags in \cyg are roughly $\propto f^{0.3}$
(i.e. time lags $\propto f^{-0.7}$), and $\Delta\phi$ is comparable to the
noise near $0.1$ Hz. As $\alpha \sim 0.5$, the weak, positive dependence of
phase lag on frequency is what allows us to detect the phase/time lags in
the $~0.1-30$ Hz range. Below $f \sim 0.1$ Hz, $\alpha \sim 0$, and
therefore $\Delta \phi/\phi \propto f^{-0.8}$, assuming $\phi \propto
f^{0.3}$.  The phase/time lags therefore become undetectable below $\approx
0.1$ Hz.  Extending our measurements to a decade lower in Fourier frequency
would require an integration time of order 40 times longer.  In Figure~9,
we present these idealized estimates of the noise limit for parameters
characteristic of our observations.

\centerline{
\epsfxsize=0.96\hsize {\epsfbox{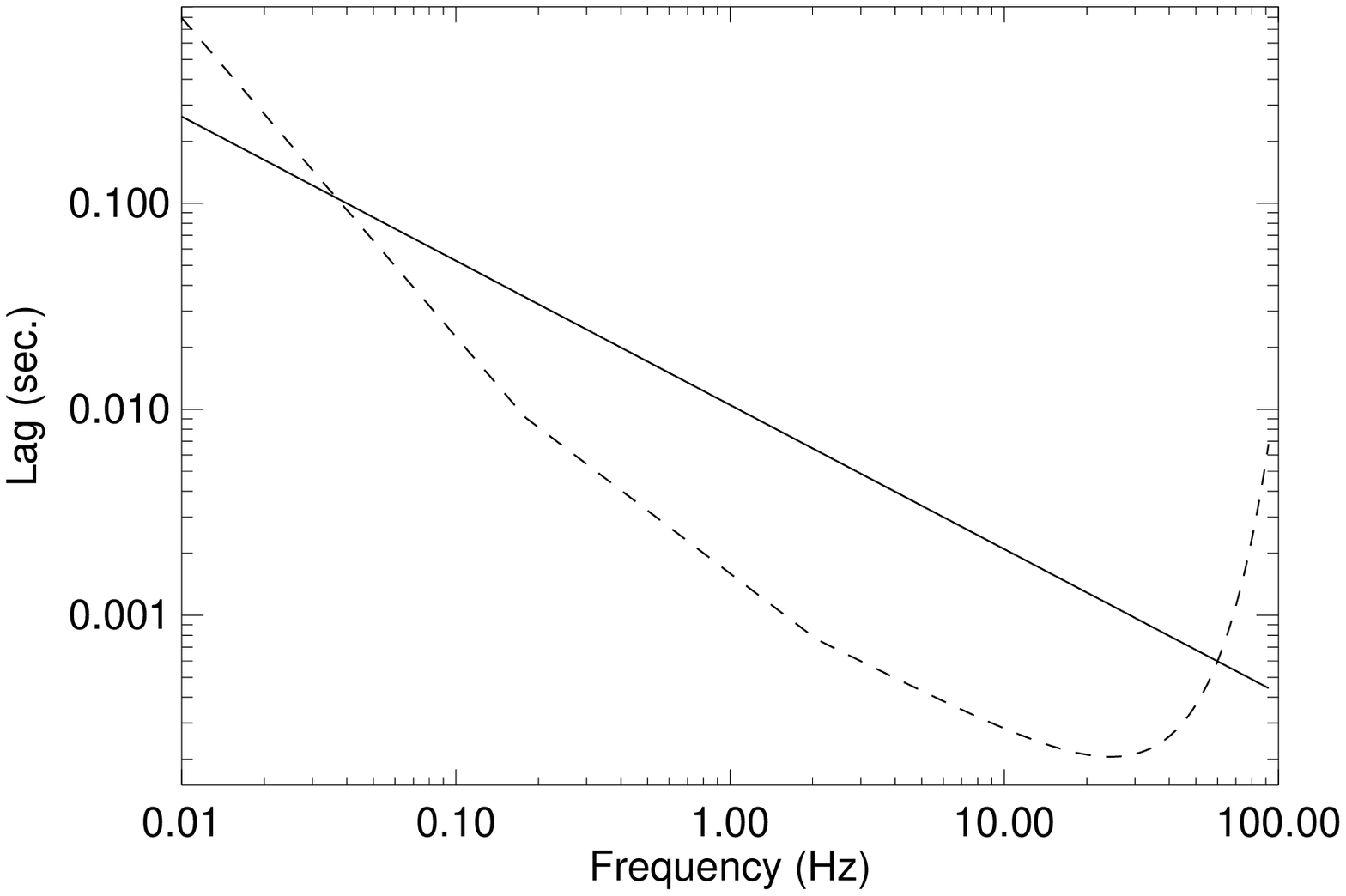}}
}
\figcaption {The dashed line is the idealized estimate of the noise level
for time lag measurements typical in this paper [cf. eq.~(\ref{eq:noise})].
We have used the double broken power law fit and count rate for the
($0$--$3.9$\,keV) energy band in our estimate, and we have taken $\gamma^2(f)
= \exp[-(f/30~{\rm Hz})^2]$.  The solid line is an idealized representation
of the time lag we see between our lowest and highest energy bands.  For
frequencies below $\sim 0.1$\,Hz and above $\sim 30$\,Hz, lags cannot be
measured because of noise limitations.}
\label{fig:noise}
\bigskip

As discussed by \citey{bendat}, we can make these noise estimates more
rigorous.  Specifically, the estimate that we use for the uncertainty in
the phase is
\begin{equation}
\Delta \phi(f) ~=~ N^{-1/2} ~ \sqrt{ {1 - g(f)} \over {2 g(f)}} ~~,
\label{eq:lagerr}
\end{equation}
where $N$, as before, is the number of lightcurves and/or frequency bins
averaged over and $g(f)$ is defined as
\begin{equation}
g(f) ~\equiv~ {{|\langle S^*(f) H(f) \rangle|^2} \over
               {\langle |S(f)|^2 \rangle} {\langle |H(f)|^2 \rangle}} ~~.
\end{equation}
In the above, $S(f)$ and $H(f)$ are the Fourier transforms of the two light
curves to be compared {\it without} Poisson noise subtracted.  This formula
has the properties discussed above (including taking account of the
coherence function, cf. \S\ref{sec:coher}), and provides a more rigorous
estimate of the errors in the phase lag than simple error propagation
estimates (e.g., \cite{cui:97b}).  In fact, $g(f)$ is simply the
\emph{measured} coherence of the two light curves, \emph{without}
correcting for counting noise.  The error in the time lag is then simply
given by $\Delta \tau(f) = \Delta \phi(f)/2 \pi f$.

\subsection{Time Lags in Cyg X-1}\label{sec:cyg_tlag}

\begin{figure}
\centerline{
\psfig{figure=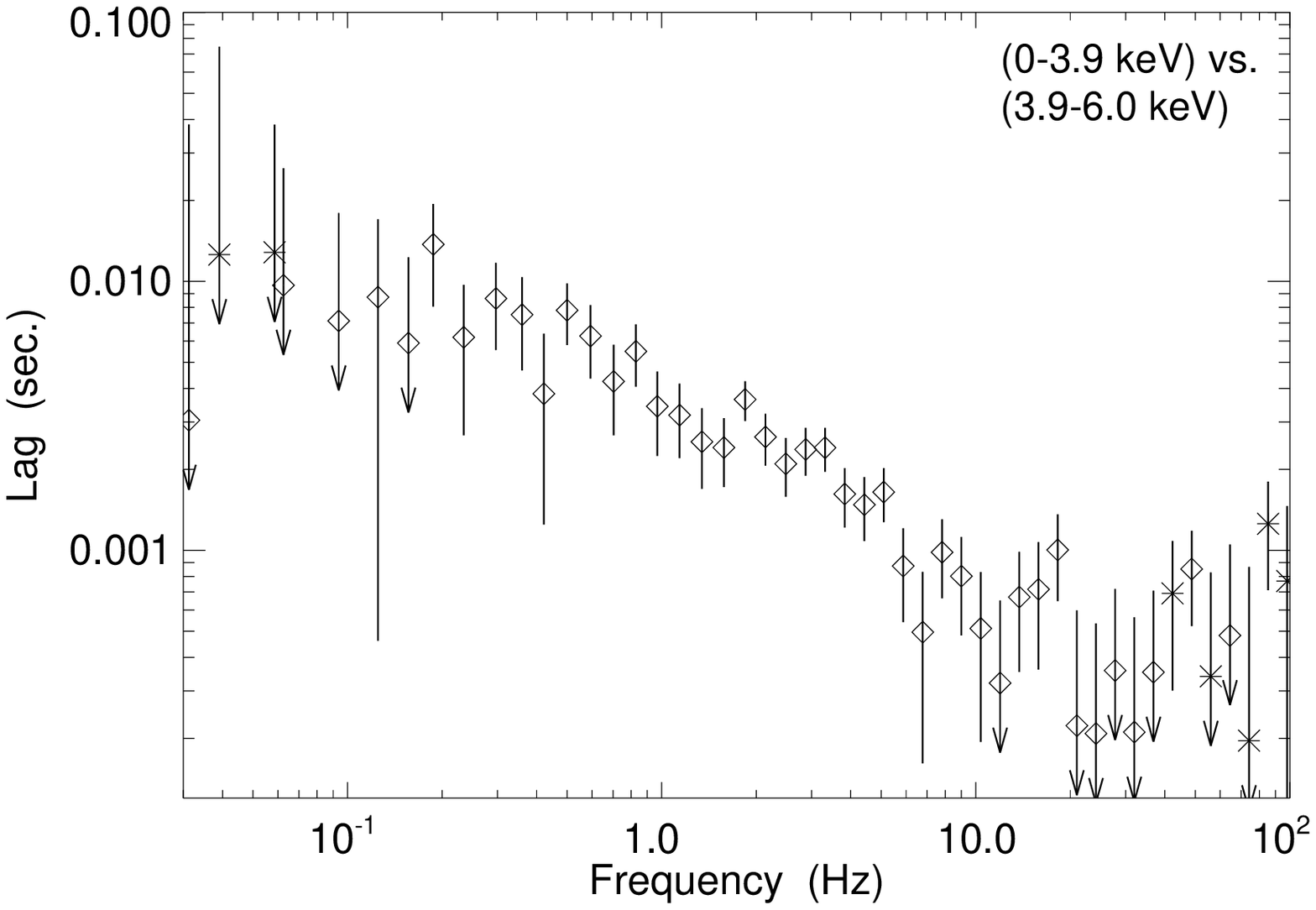,width=0.456\textwidth}
\psfig{figure=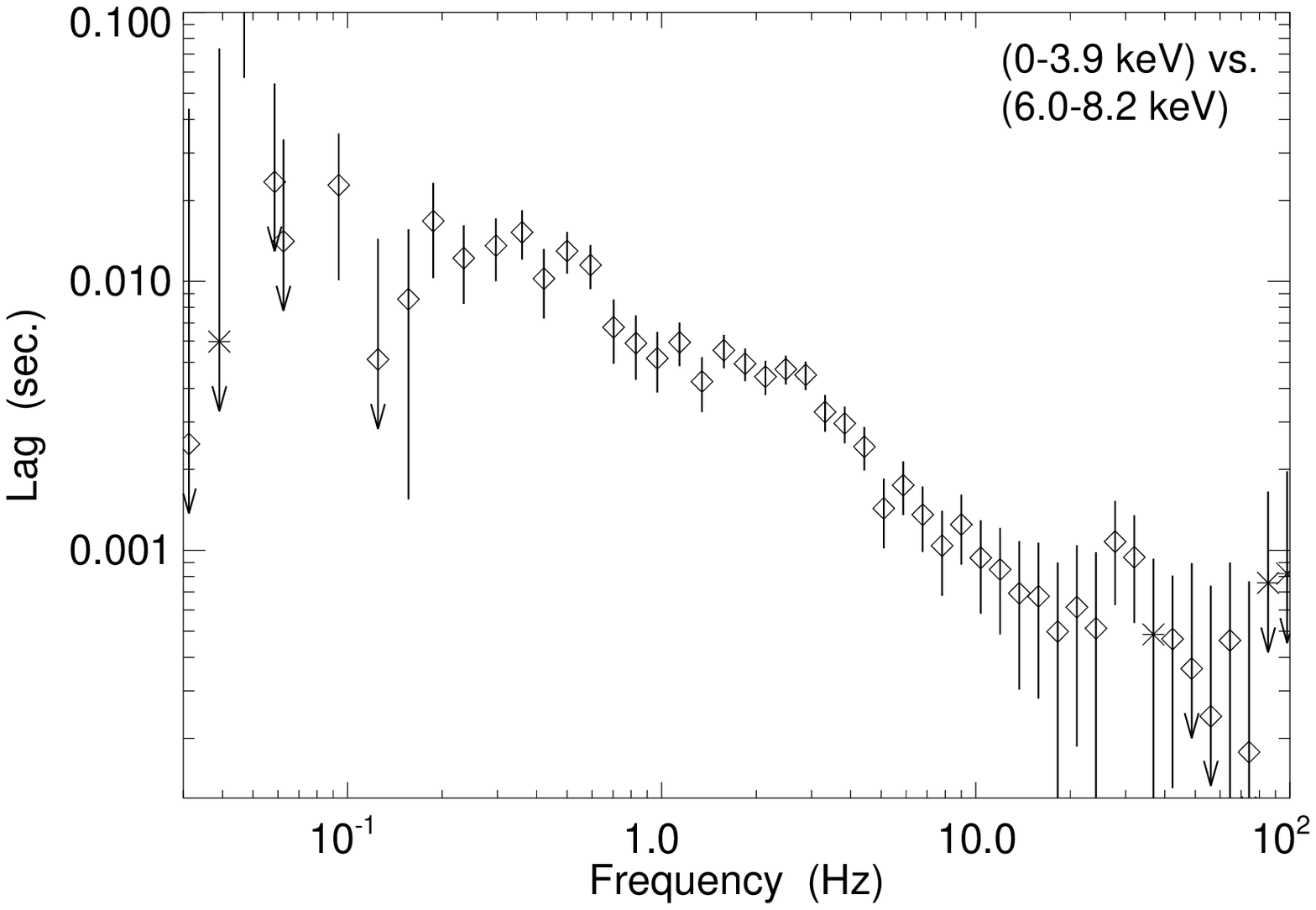,width=0.456\textwidth}
}
\centerline{
\psfig{figure=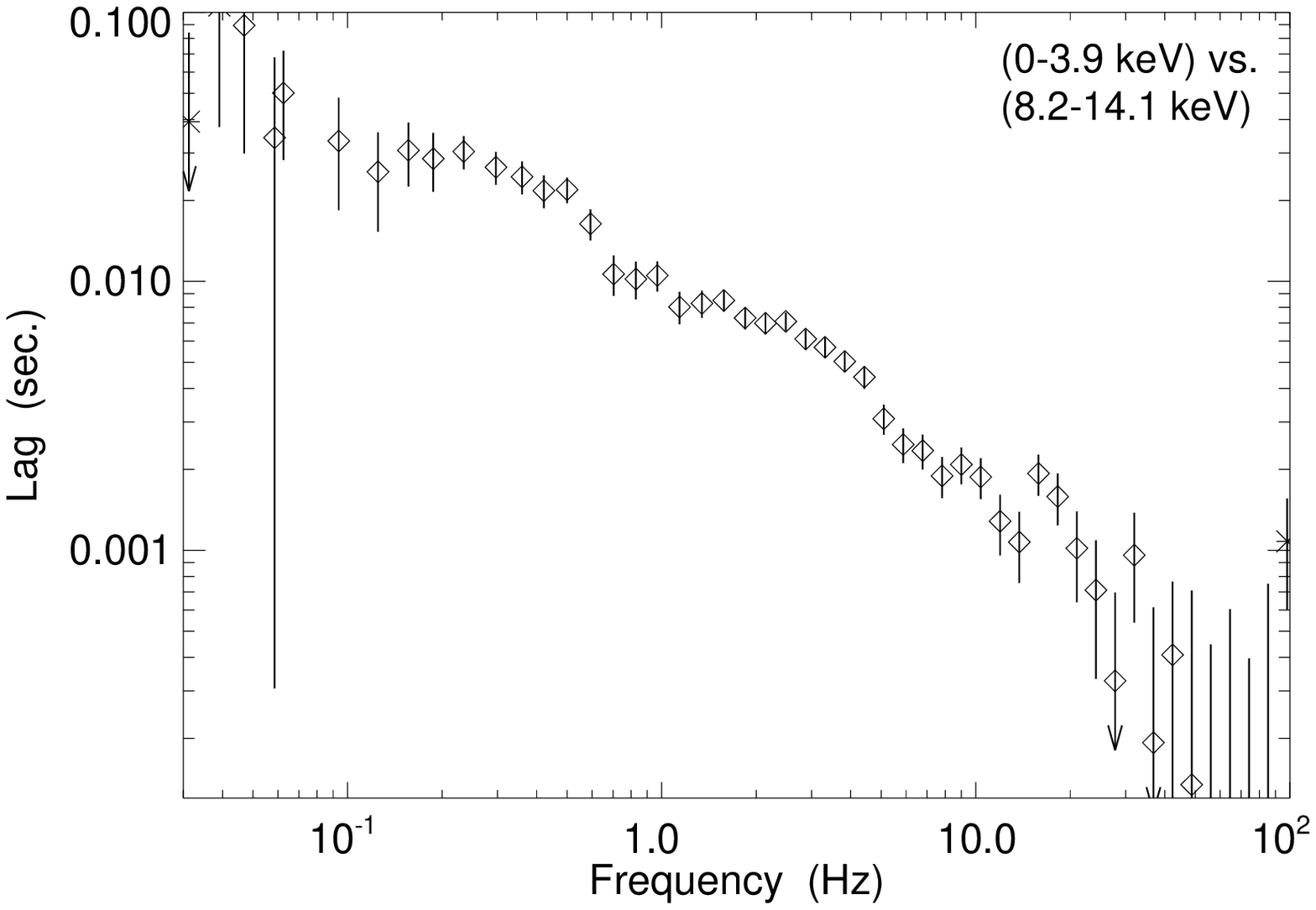,width=0.456\textwidth}
\psfig{figure=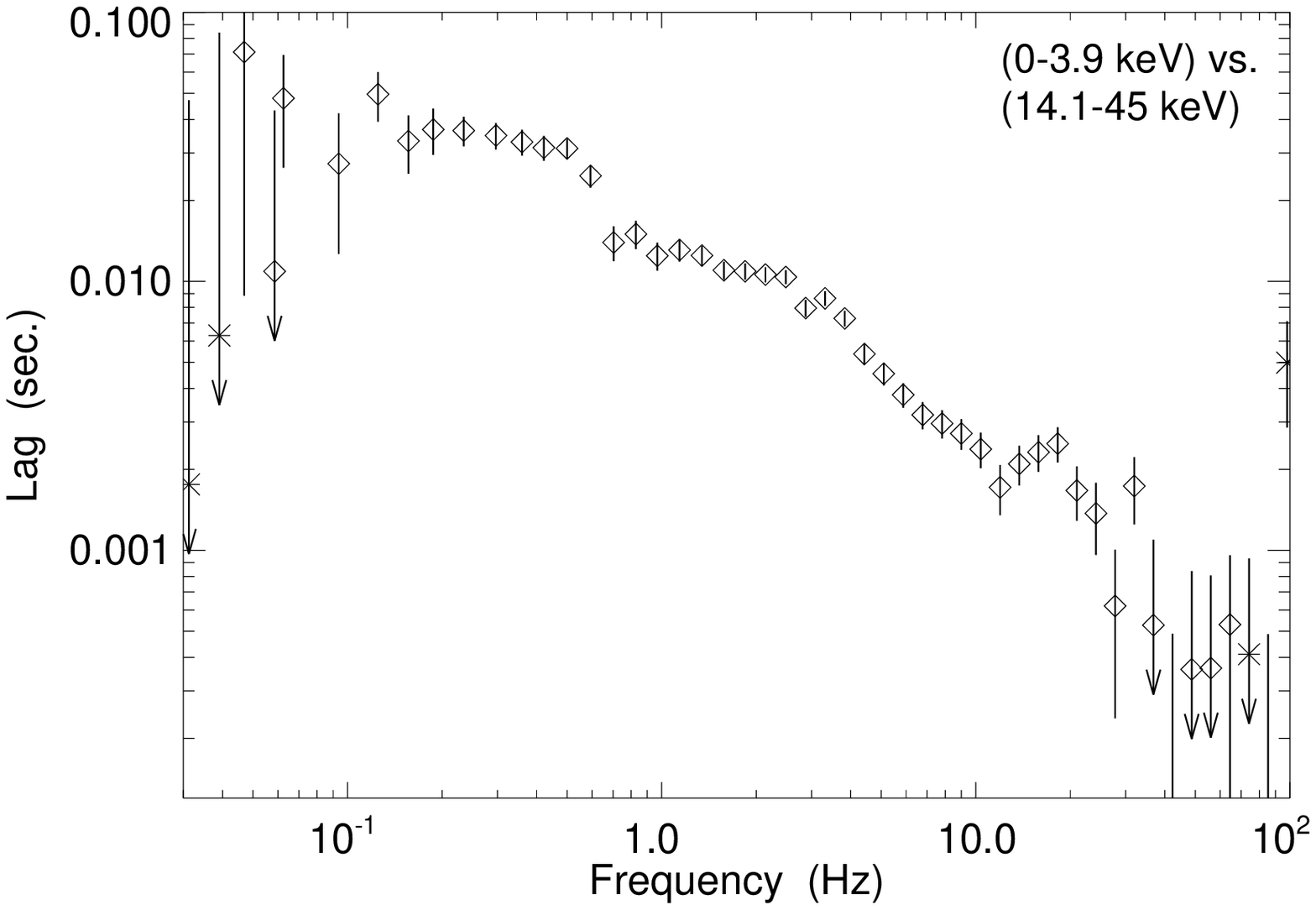,width=0.456\textwidth}
}
\caption
{Time lags, as a function of Fourier frequency, for various energy bands
vs. the lowest energy band (0$-$3.9 keV).  Diamonds represent hard lagging
the soft, whereas asterisks represent soft lagging the hard.  Error bars
are calculated from eq. \ref{eq:lagerr}.  Measurements are above the noise
level in roughly the range of $0.1$--$30$ Hz.}
\label{fig:lags}
\end{figure}

In Figure~\ref{fig:lags}, we present the time lags vs. Fourier frequency for
energy bands $2$--$5$ vs. energy band $1$.  The measurements are above the
noise level, as defined by eq.~(\ref{eq:lagerr}), from $0.1$--$30$ Hz.  Within
this frequency range, the data are consistent with the softest energy band
{\it always} leading the harder energy bands.  Those frequency points where
the hard leads the soft are dominated by Poisson noise fluctuations.

Some general properties of the observations are notable.  First, the time
lags between all channels approximately show a power law dependence upon
Fourier frequency, $\tau(f) \propto f^{-0.7}$, although there are
significant deviations from a simple power law.  This is consistent with
previous observations of the low state of \cyg
(\cite{miyamoto:89a,miyamoto:92a,crary:98a}).  As discussed in
\S\ref{sec:tlsimp}, such a frequency dependence is closer to a constant
phase lag (as a function of Fourier frequency) than a constant time lag.
The time lags presented in Fig.~\ref{fig:lags} correspond to Fourier phases
ranging from 0.01 (low Fourier frequency, energy band 2 vs. energy band 1)
to 0.2 radians (high Fourier frequency, energy band 5 vs. energy band 1).
In terms of a constant phase lag transfer function interpretation
[cf. eq.~(\ref{eq:cptrans})], this means that nearly all of the hard and soft
flux variations occur quasi-simultaneously, with only a small fraction of the
hard photons lagging the soft photons.

The $f^{-0.7}$ dependence of the time lag also implies a large dynamic
range in delay times, on the order of a decade and a half.  The longest
time lags are 0.05 sec. Measured as a light crossing time, this corresponds
to a distance of $10^3~GM/c^2$, for $M=10~M_\odot$.  Between energy band 5
and energy band 1, the shortest measured time lag is $\approx 10^{-3}$ sec,
which corresponds to a light crossing radius of $20~GM/c^2$ for
$M=10~M_\odot$.  It is this large dynamic range in characteristic length
scales associated with the variability, as further discussed in
\S\ref{sec:whatitmeans}, that is difficult for theoretical models to
address (although see \cite{kazanas:97a}).  However, if one were to invoke
slower ``propagation speeds'' (see also paper III), the implied length
scales would be commensurately shorter.  If one were to restrict the size
scale to $50~GM/c^2$, the longest time lags would imply propagation speeds
of $0.05~c$.  As discussed in paper III, this is uncomfortably slow for
both Advection Dominated Accretion Flow (ADAF) models, where one expects
the matter in the inner regions to be nearly in radial free fall
(cf. \cite{narayan:96e}), \emph{and} Compton corona models, where the sound
speed of the corona is a sizable fraction of $c$.

\end{multicols}
\section*{}
\vskip -15pt

The time lags generally increase with energy band separation. We
quantify this statement in \S\ref{sec:whatitmeans} below. This increase
can be reconciled with either the notion that ``propagation path'' from
soft source to hard response increases with energy separation, or that the
``propagation speed'' from soft source to hard response decreases with
energy separation.  We also note that the time lags between bands 1 and 5
are consistent with the sum of the time lags from bands 1 to 2,
bands 2 to 3, bands 3 to 4, and bands 4 to 5.

\section{Physical Interpretations}\label{sec:whatitmeans}

\subsection{PSD Shape}\label{sec:psdshape}

In paper III, we consider in detail the implications of our timing analysis
to the specific Comptonization model that we presented in paper I.  We also
consider in more general terms the implications of our timing analysis for
ADAF models.  Specifically, we investigate the phenomenology of what we dub
``propagation models'', and derive constraints on likely propagation speeds
for disturbances in a disk.  Here, however, we adopt a slightly more
general approach.

We still wish to consider the implications of Comptonization models;
however, rather than discuss a specific model, we will consider those
effects that we generally expect to find.  As we have discussed elsewhere
(\cite{nowak:96a}, and references therein), if the source of the observed
variability is from soft photons, perhaps from an accretion disk, that are
subsequently upscattered by a Comptonizing medium, then we expect a number
of effects to be evident.

First, at high Fourier frequencies, we expect that dispersion in the
scattering times in the Comptonizing medium will lead to an attenuation in
the variability amplitude.  This attenuation will be manifested as a
(typically exponential) rollover in the PSD on timescales shorter than the
typical diffusion timescale through the Comptonizing medium
(\cite{brainerd:87a,kylafis:87a,wijers:87a,stollman:87a,bussard:88a,kylafis:89a,miller:92a,miller:95a,nowak:96a}).
As discussed in \citey{nowak:96a}, the {\it simplest} expectation is an
exponential rollover in the PSD that is $\propto \exp[-(f\tau)^2]$, where
$\tau \sim 2 \pi \sqrt{N} \bar \lambda_{\rm es}/c$.  Here, $N$ is the number of
scatters that the photon has undergone, and $\bar \lambda_{\rm es}$ is the mean
free path between scatters.  If the optical depth of the Comptonizing
medium is $\approx 1$, then $\bar \lambda_{\rm es} \approx R$, where $R$ is the
size of the medium.

In the simplest models of Comptonization (cf. \cite{pozd:83a}, and
references therein), one expects the number of scatters $N$ to be of the
order
\begin{equation}
N \sim {{mc^2}\over{4kT}} ~\ln \left( {{E_o}\over{E_s}} \right ) ~~,
\label{eq:nscat}
\end{equation}
where $T$ is the temperature of the Comptonizing medium, $E_s$ is the
energy of the soft input photons, and $E_o$ is the energy of the observed
output photons.  For a very simple Compton cloud model, one therefore
expects an exponential cutoff frequency, $f_c \equiv \tau^{-1}$, on the
order of
\begin{equation}
f_c \approx 50 ~{\rm Hz}  {{\left( {{R}/{50~GM/c^2}} \right )^{-1} ~
\left ( {{kT}/{80~{\rm keV}}} \right )^{1/2}} \over {\left [
\ln \left (  {{E_o}/{1~{\rm keV}}} \right ) - \ln \left ( {{E_s}/
{0.2~{\rm keV}}} \right ) + 1.6  \right ]^{1/2} }} ~,
\label{eq:cutfreq}
\end{equation}
where we have taken $M = 10~M_\odot$ and where we have chosen a characteristic
coronal radius and temperature consistent with the spectral model of paper~I.

In Figure~11 
we show the results of fitting an exponentially cutoff power law to the
PSDs of \S\ref{sec:psd}.  For all the energy bands, we fit functions of
the form $a f^b \exp[-(f/f_c)^2]$ to the $3-90$ Hz data.  The reduced
$\chi^2$ for the fits ranged from $1.8$ for the lowest energy band to
$5.2$ for the highest energy band. Again, reduced $\chi^2$ as large as
these are partly indicative of the extremely small error bars associated
with the data, as well as the fact that a cutoff powerlaw is only a
qualitative description of the data.  As also discussed in \S\ref{sec:psd}
and shown in Figure~11b,
the PSD slope ``hardens'' as one goes to higher energy bands.  The best
fit slopes ranged from $-1.74$ for the lowest energy band to $-1.4$ for
the highest energy band.  The change in slope is significant and appears
to be proportional to the $\log$ of the band energy.

Although exponential cutoffs are allowed for all the energy bands, fitting
pure power laws (as was done in \S\ref{sec:psd}) results in nearly equally
good fits. Based upon eq.~(\ref{eq:cutfreq}), we would have expected much
stronger cutoffs, {\it if the source of the high frequency variability is
intrinsic to the soft input photons}\footnote{Note that this statement is
true for a relatively spatially compact source.  In the model of
\citey{kazanas:97a}, the {\it entire} PSD is reprocessed white noise from a
soft input source.  The range of power law slopes is achieved by
distributing the Compton scatterings over several decades in radius of the
system.  We elaborate upon this point further in \S\ref{sec:lagimps}.}.  We
have fit a function of the form of eq.~(\ref{eq:cutfreq}) to our best fit
cutoff frequencies, which we present in Figure~11c.
The best fit numerator was $123$ Hz [i.e. $(R/50~GM/c^2)^{-1}
(kT/80~keV)^{1/2} \approx 2.5$], and the best fit soft input energy was
$E_s = 2~{\rm keV}$.  We emphasize that we {\it do not} consider this a
detection of an exponential rollover, but more of an upper limit to the
degree of rollover (or equivalently a lower limit to the cutoff frequency,
$f_c$).  In Figure~11c
we present the cutoff frequencies that we naively would have expected if
$E_s = 0.2~{\rm keV}$ (dashed line); however, we have kept the numerator as
$123$ Hz.  The expected value of the exponential cutoff frequencies are
somewhat lower than the values permitted by the data. (We further discuss this
point in paper~III.)

The simplest explanation (although see \cite{kazanas:97a} and
\S\ref{sec:lagimps} below) for the increase in the PSD ``hardness'' with
energy and for the lack of a strong rollover in the high energy PSD is that
the high frequencies {\it do not} represent variability intrinsic to the
soft input source, but rather are the result of {\it direct} modulation by
fluctuations in the Comptonizing medium itself.  As we expect the
Comptonizing medium to be both very hot and concentrated near the compact
object where the dynamical timescales are the shortest (cf. paper I), it is
natural to expect the corona to show harder PSD slopes at higher energies.

We hypothesize that the PSD above the break at $\approx 2$\,Hz is dominated
by fluctuations of the corona on dynamical timescales.  As discussed in
\citey{vaughan:97a} and in \S\ref{sec:cohimp} below, this is also
qualitatively consistent with the loss of coherence seen at these high
frequencies.  We further consider the characteristic PSD frequencies below.

\centerline{ 
\epsfxsize=0.96\hsize {\epsfbox{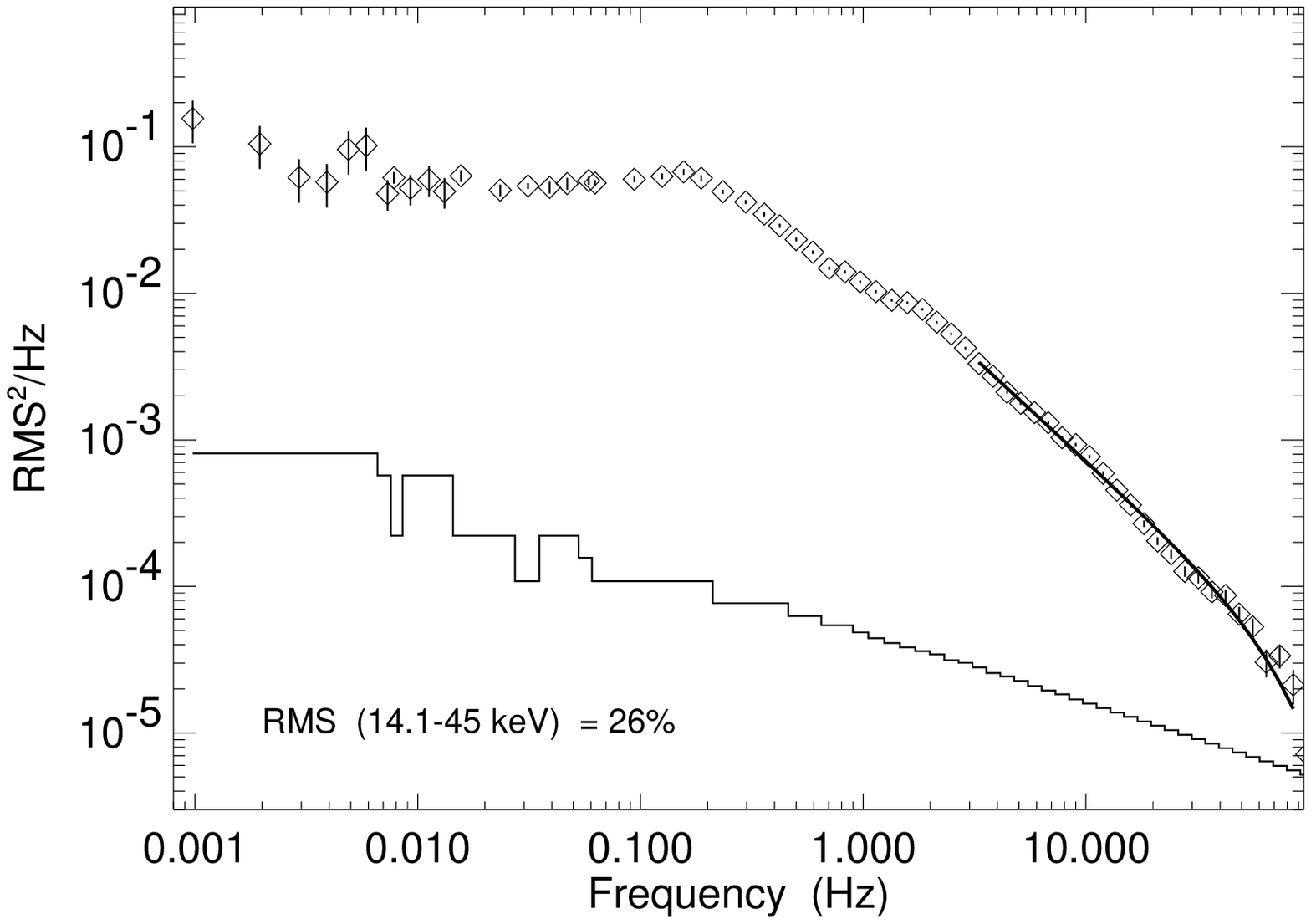}} }
\centerline{ 
\epsfxsize=0.96\hsize {\epsfbox{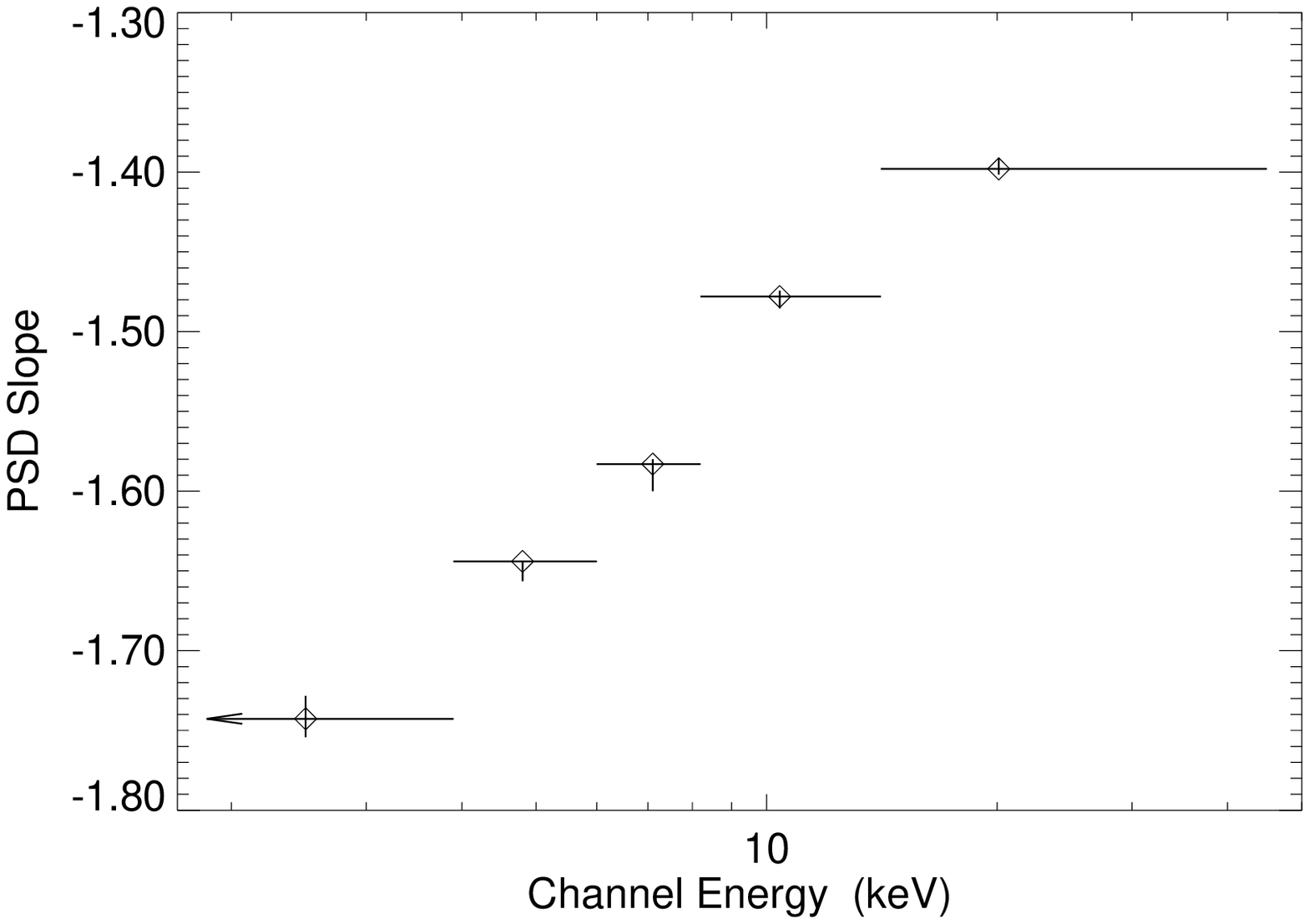}} }
\centerline{ 
\epsfxsize=0.96\hsize {\epsfbox{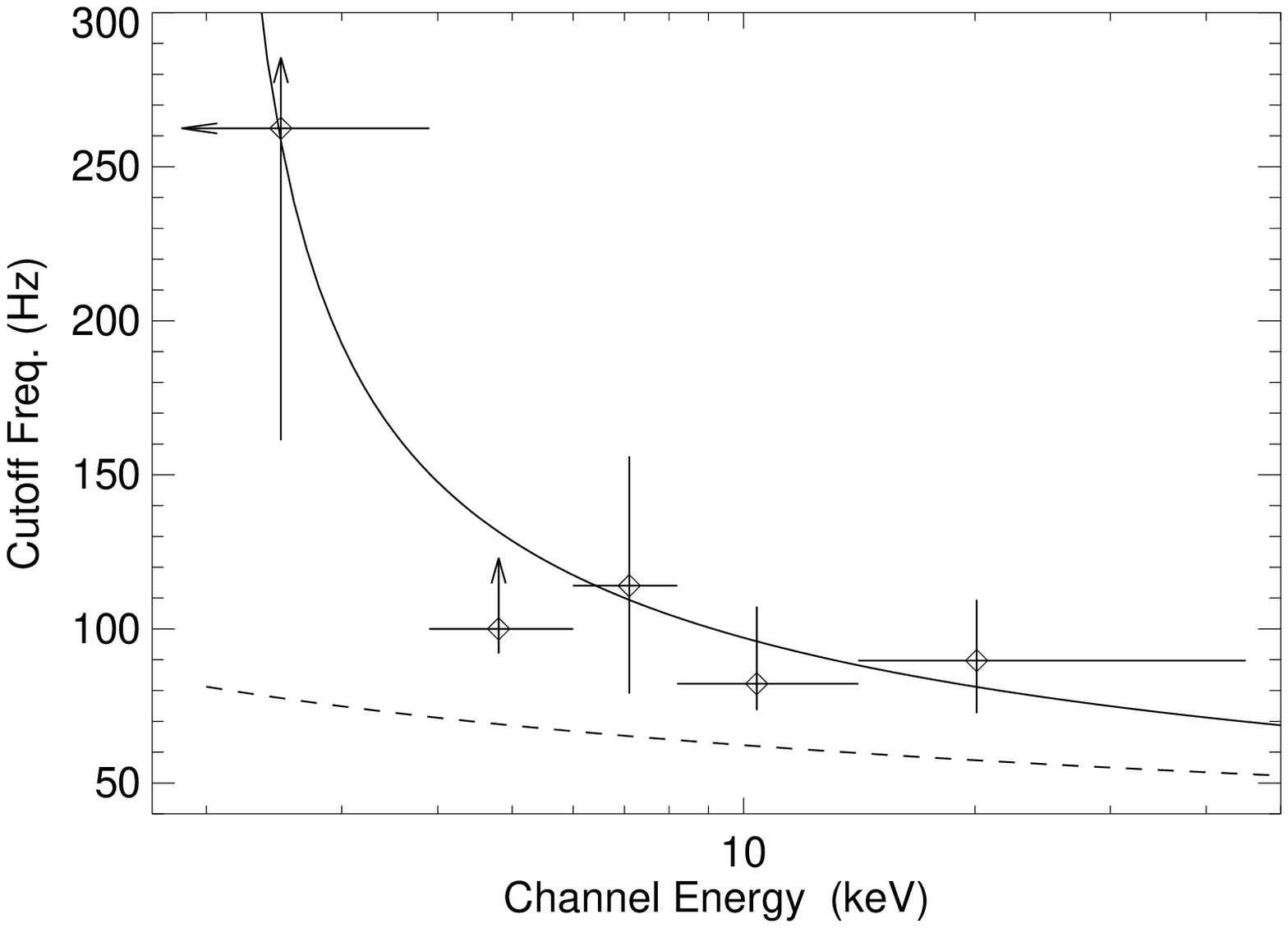}} }
\smallskip
\figcaption {{\sl Top:} The same energy band 5 PSD as presented in
Fig.~\ref{fig:psds} with an exponentially cutoff power law fit to the
$3$--$90$\,Hz data.  {\sl Middle:} Best fit power slopes to the
$3$--$90$\,Hz PSD data as a function of channel energy.  {\sl Bottom:} Best
fit cutoff frequency (solid line), using eq.~(\ref{eq:cutfreq}). The best
fit numerator is $123$\,Hz with $E_s = 2$\,keV.  For the dashed line we
have fixed the numerator of eq.~(\ref{eq:cutfreq}) to $123$\,Hz, but
have taken $E_s = 0.2$\,keV.}
\label{fig:expcuts}
\bigskip

\vfill\eject

\subsection{Characteristic Timescales of the PSD}\label{sec:chartimes}

The Compton corona model that we presented in paper I consisted of a
spherical, central hot cloud surrounded by a thin, cold disk.  The
temperature at the inner edge of the cold disk was taken to be 200 eV, and
there was evidence for an iron line with equivalent width $\approx 40$ eV
(cf. \S\ref{sec:spectra}).  These latter two facts, considering that the
model was reasonably consistent with the data, suggest a radius for the
Compton corona of approximately $50~GM/c^2$ if
$M=10~M_\odot$. Substantially smaller radii would imply a hotter
temperature on the inner edge of the disk, which is not consistent with the
data (\cite{dove:98a}).  Substantially larger radii would imply very weak
iron lines, unless a substantial fraction of the coronal energy release
were also at large radii, as in the model of \citey{kazanas:97a}.

Taking $50~GM/c^2$ as a characteristic radius, we have the following
characteristic timescales for a $10~M_\odot$ black hole accreting at
approximately $5\%~L_{\rm Edd}$ (cf. Table~\ref{tab:cygpar}). The light
crossing timescale is of order $2.5\times10^{-3}~{\rm sec}$, which is of
order the shortest Fourier time lags seen in the system.  The dynamical
timescale (i.e., the Keplerian orbital period) is $0.1$\,s, or an orbital
frequency of $9$\,Hz.  This is of the same order as  the location
of the break in the PSD at $\approx 2$\,Hz, and is also consistent with
the Fourier frequencies above which the soft and hard X-ray become
incoherent with each other.  This further supports our hypothesis that the
variability above $\approx 2$ Hz is dominated by {\it direct} modulation of
the corona at radii $\aproxlt 50~GM/c^2$.  We again note that there is no
evidence for any energy dependence of the PSD slopes below 2\,Hz.

The characteristic viscous timescale is of order the dynamical timescale
divided by the Shakura-Sunyaev $\alpha$-parameter and by the square of the
fractional Eddington luminosity (\cite{shakura:73a,shakura:76a}).  Thus
this timescale is of ${\cal O}(50-500~{\rm s})$, or equivalently ${\cal
O}(2 \times 10^{-3}$--$2 \times 10^{-2}~{\rm Hz})$, for $\alpha \sim
0.1$--$1$.  Characteristic thermal timescales are of the order of the
dynamical timescale times the $\alpha$-parameter, i.e. $1$\,s for $\alpha
\approx 0.1$.  The break in the PSDs at $\approx 0.2$\,Hz could therefore be
associated with a characteristic viscous timescale or thermal timescale.
The PSD $\propto f^{-1}$ between $0.2$--$2$\,Hz could be indicative of viscous
and/or thermal fluctuations over a range of radii $\aproxlt 50~GM/c^2$.
Alternatively, one might wish to associate this portion of the PSD with
dynamical timescale fluctuations at radii $\aproxgt 50~GM/c^2$.

The feature at $\approx 5\times10^{-3}$\,Hz is somewhat problematic.  This
frequency is at the lower end of what we would associate with a
characteristic viscous timescale (unless $\alpha$ is even smaller than
$0.1$), and is too slow for either a thermal or dynamical timescale unless
it is associated with a very large radius ($R \approx 3 \times 10^{4}
~GM/c^2$ for a dynamical timescale).  We again stress, however, that based
upon the PSDs alone, the evidence for a low frequency QPO is very weak.

\subsection{Implications of the Time Lags}\label{sec:lagimps}

The time lags seen in \cyg are often interpreted in terms of Comptonization
models (cf. \cite{cui:97b,kazanas:97a,crary:98a} and references
therein). Taking the simplest Comptonization models, one naturally expects
the hard photons to lag behind the soft photons as the hard photons have
undergone more scatters.  The expected time delay is therefore proportional
the difference in the average number of scatters required to reach the hard
energy channel compared to the soft energy channel times the average
distance traveled between scatters. Again taking the electron scattering
optical depth $\approx 1$ so that $R \approx \bar \lambda_{\rm es}$, the
expected time delay is of the order
\begin{equation}
\Delta \tau ~\approx~ 1.6 ~ \left ( {{R}\over{50~GM/c^2}} \right )
      \left( {{kT}\over{80~{\rm keV}}} \right )
\ln \left ( {{E_H}\over{E_S}} \right )~{\rm ms} ~~,
\label{eq:loge}
\end{equation}
where $E_H$ is the energy of the hard energy channel and $E_S$ is the
energy of the soft energy channel. (As before, we have taken $M=10~M_\odot$.)

\centerline{ 
\epsfxsize=0.96\hsize {\epsfbox{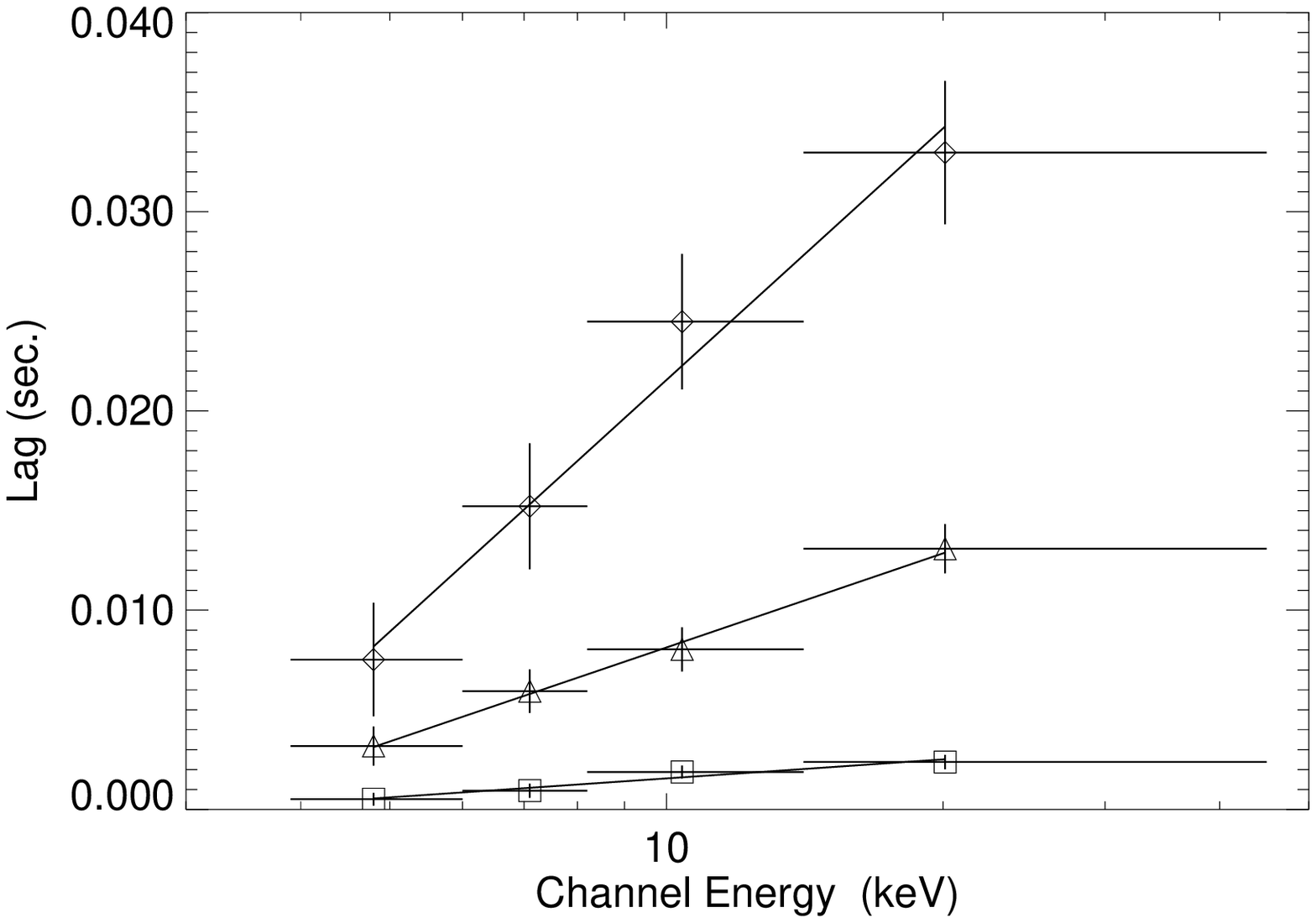}} }
\smallskip
\figcaption
{Time lags as a function of energy compared to the lowest energy band,
($0-3.9$)\,keV, for three different Fourier frequencies.  Diamonds are for
$0.3$\,Hz, triangles are for $1$\,Hz, and squares are for $10$\,Hz.
Lines are the best fit functions of the form of eq.~(\ref{eq:loge}).}
\label{fig:loge}
\bigskip

In Figure~12
we present the time delay, compared to the lowest energy band, as a
function of energy for three frequencies, $0.3$\,Hz, $1$\,Hz, and
$10$\,Hz. As has been noted recently for other observations
(\cite{cui:97b}), the time delays do indeed increase nearly linearly with
the $\log$ of the band energy.  We have fit a function of the form of
eq.~(\ref{eq:loge}) to these data, and find the best fit intercept to be
given for $E_S \approx 3.0-3.2$\,keV, in reasonable agreement with the
average energy of $2.5$\,keV for the lowest energy channel.  (We {\it did
not} include a 0 sec time delay between the lowest energy band and itself
in the fits, and the y-intercept was a free parameter.)  Other authors
(e.g. \cite{cui:97b,crary:98a}) recently have claimed that this previously
observed logarithmic energy dependence of the time lag is strong evidence
for the presence of a Compton corona; however, rarely do they note the fact
that although the functional form agrees, the best fit slopes are
apparently functions of Fourier frequency.

Specifically, if we set $kT = 80~{\rm keV}$ in eq.~(\ref{eq:loge}),
then we find best fit values of $R=230~GM/c^2$ for $f=0.3$\,Hz,
$R=85~GM/c^2$ for $f=1$\,Hz, and $R=17~GM/c^2$ for $f=10$\,Hz.  Any
individual Fourier frequency agrees well with the simple expectations of
Comptonization theory; however, taken as a whole the time delays {\it do
not} agree with the {\it simplest} Comptonization models.  This fact was
originally noted by \citey{miyamoto:89a}.  Two possibilities have been
suggested for maintaining the logarithmic energy dependence, yet creating
the large dynamical range in time delays, within the context of
Comptonization.  The first is the possibility that the time delays are
inherent to the source of soft input photons and are merely ``reprocessed''
by the Comptonizing medium (\cite{miller:95a,nowak:96a}).  We explore this
possibility in detail in paper III.

The second suggestion is that the dynamic range in time delays reflects a
large range in radii of a Comptonizing medium (\cite{kazanas:97a}).
Specifically, \citey{kazanas:97a} posit a uniform temperature corona with
essentially equal optical depth per decade of radius (i.e., electron density
approximately $\propto r^{-1}$) that extends for roughly three decades in
radius.  A white noise (i.e., flat PSD) input spectrum is reprocessed into a
power law PSD that shows time delays comparable to those shown in
Figure~12. 
Specifically, the time delays, at a given Fourier frequency, show a
logarithmic energy dependence, and higher Fourier frequencies show shorter
time delays.  This latter attribute is due to the fact that any portion of
the lightcurve that comes from photons that have scattered over large
distances has its high frequency variability strongly attenuated
[cf. eq.~(\ref{eq:cutfreq})].  Any observed high frequency variability is
{\it only} sampling those photons that have scattered over short distances,
and therefore it will show relatively short time delays between soft and
hard photons. Furthermore, the $r^{-1}$ density profile means that there is
roughly equal probability of scattering on length scales in any given
decade of radius, which leads to the observed dynamic range in the time
delays.  Although the model of \citey{kazanas:97a} is consistent with the
data, it is difficult to understand how such a configuration would evolve.
It implies that the integrated electron energy goes as $r^2$, i.e. most of
the electron thermal energy is at the largest radii.  For the Compton
corona model we discuss in paper I, the bulk of the accretion and electron
thermal energy is expected to be at radii $\aproxlt 50~GM/c^2$.

\subsection{Implications of the Coherence}\label{sec:cohimp}

The physical implications of the coherence function are discussed
extensively in \citey{vaughan:97a}. Here we point out some of the specific
implications for Comptonization models.  If the low frequency variability
is reprocessed from the soft input photons, the near unity coherence
between $\approx 0.02-10$\,Hz indicates that on these timescales the
Comptonizing medium must appear as a {\it static} medium.  [As discussed by
\citey{nowak:96a}, a static corona is a {\it linear} reprocessor.]
If, on the other hand, the corona directly creates the low frequency
variability, then it must do so in a global, linear fashion, despite the
fairly large observed variability amplitudes.

As discussed by \citey{vaughan:97a}, the loss of high frequency coherence
is consistent with the presence of multiple, independent sources and
responses.  As the frequencies associated with the dynamical timescales are
$\aproxgt 9$\,Hz for $R\aproxlt 50~GM/c^2$ and $M=10~M_\odot$, the high
frequency coherence loss and the ``hardening'' of the high frequency PSD
with energy is consistent with the hypothesis that we are directly viewing
the formation timescales and processes of the corona.

The loss of coherence below $\aproxlt 0.02$\,Hz is consistent with viscous
timescales for $R \aproxgt 50~GM/c^2$.  Perhaps on these low frequency
timescales we are seeing the disk `global' structure becoming
disassociated from the corona `global' structure.

We again note that although there seems to be a general trend for the
coherence to drop for frequencies $\aproxlt 0.02$\,Hz, there apparently is
a recovery at $0.005$\,Hz, coincident with the weak feature in the PSD.  In
general, we expect any global oscillation to be fairly coherent.  The
coherence associated with this feature makes it appear to be more
significant than implied by the rather low $50$--$85\%$ value suggested by
a Lomb-Scargle periodogram (cf. \S\ref{sec:psd}).  Unfortunately, we are
not aware of any direct way of assessing the statistical significance of
this coherence peak coincident with the weak PSD feature.

\section{Summary}

We have presented timing analysis for a 20\,ks RXTE observation of
{Cygnus~X--1}, taken approximately two weeks after \cyg returned to its
hard/low state from its soft/high state.  It is possible that some of the
properties of \cyg were slightly different from their ``standard'' hard/low
state values (e.g. the $10$--$200$\,keV photon index).  However, many of
the timing properties were similar to previous observations
(\cite{miyamoto:89a,belloni:90a,belloni:90b,miyamoto:92a}).

We have presented detailed discussions of all the timing analyses used in
this work, including discussions of errors and noise limits.  We are very
confident of our results for the PSD from $10^{-3}$--$100$\,Hz, for the
coherence from $10^{-3}$--$30$\,Hz, and for the time delay from
$0.1$--$30$\,Hz.  The coherence function is the most sensitive to
systematic errors (i.e., deadtime); however, the general trend of coherence
loss out to $\approx 100$\,Hz is secure even with uncertainties as large as
$25\%$ in the deadtime.  ({\it Ginga} systematic uncertainties were such
that one could not be confident in coherence trends above $\approx 1$\,Hz;
cf. \cite{vaughan:91a,vaughan:97a}.)

Some of the notable features of our observations were the $0.005$\,Hz
feature and the high frequency `hardening' with energy seen in the PSD.
The former, if real, could be indicative of an oscillation on viscous
timescales, whereas the latter may be indicative of dynamical timescale
variability produced directly by the corona.  The coherence function lends
support to both of these interpretations.  For the former, the coherence is
seen to recover towards unity, whereas for the latter, the coherence is
seen to drop in accord with theoretical expectations of multiple flaring
regions (\cite{vaughan:97a}).  The coherence was also seen to be nearly
unity from $0.02$--$10$\,Hz.  Curiously, the $0.01$--$0.2$\,Hz coherence
between the lowest and highest energy bands was seen to be slightly greater
than the coherence between the lowest and second highest energy bands.

We also made comparisons of our timing analysis to the expectations of
simple Comptonization models.  The high frequency PSD in the highest energy
channels was {\it not} as attenuated as expected for simple corona models,
again lending credence to the hypothesis that we are seeing direct
modulation by a fluctuating corona.  Although the time lags show the
expected logarithmic dependence upon energy, their increase with 
decreasing Fourier frequency is contrary to the simplest expectations for a
relatively compact, uniform corona.  We explore this point further in a
companion paper (paper III).

Both our energy spectral analysis of paper I and the timing analysis
presented here show that the simplest models are inadequate for describing
the high quality RXTE data. Due to its broad spectral coverage, RXTE allows
us to compare and discriminate among detailed, self-consistent models for
the energy spectrum.  Coupling these detailed models (as we do in paper
III) with RXTE's superior timing capabilities places further constraints on
their structure, and more importantly, on their dynamics.  Only those
models in which {\it both} the spectral and temporal data can be explained
should be considered valid candidates for {Cygnus~X--1}.

\acknowledgements 

\vskip-0.2 true in

We would like to acknowledge useful discussions with P. Michelson,
E. Morgan, K. Pottschmidt, R. Staubert, M. van der Klis, and W. Zhang.
This work has been financed by NSF grants {AST91-20599},
{AST95-29175}, {INT95-13899}, NASA Grants {NAG5-2026}, {NAG5-3225},
{NAGS-3310}, DARA grant {50 OR 92054}, and by a travel grant to
J.W. from the DAAD.


\begin{thebibliography}{}

\bibitem[\protect\astroncite{Aab et~al.}{1984}]{aab:84a}
Aab, O.~{\'E}., Bychkova, L.~V., Kopylov, I.~M., \& {Kuma\u{\i}gorodskaya},
  R.~N.,  1984, Sov. Astron., 28, 90 (orig.: AZh, 61, 152, 1984)

\bibitem[\protect\astroncite{Angelini \& White}{1992}]{angelini:92a}
Angelini, L., \& White, N.~E.,  1992,
\newblock IAU Circ.~5580

\bibitem[\protect\astroncite{Angelini, White \& Stella}{1994}]{angelini:94a}
Angelini, L., White, N.~E., \& Stella, L.,  1994,
\newblock in New Horizon of {X}-Ray Astronomy, ed. F. Makino, T. Ohashi,
  (Tokyo: Universial Academy Press),  429

\bibitem[\protect\astroncite{{Ba\-\l{}u\-ci\'nska-Church}
  et~al.}{1995}]{balu:95a}
{Ba\-\l{}u\-ci\'nska-Church}, M., Belloni, T., Church, M.~J., \& Hasinger, G.,
  1995, A\&A, 302, L5

\bibitem[\protect\astroncite{Balog, {Gon\-char\-ski\u\i} \&
  {Che\-re\-pash\-chuk}}{1981}]{balog:81a}
Balog, N.~I., {Gon\-char\-ski\u\i}, A.~V., \& {Che\-re\-pash\-chuk}, A.~M.,
  1981, Sov. Astron. Lett., 7, 336 (orig.: Pis'ma Astron. Zh., 7, 605, 1981)

\bibitem[\protect\astroncite{Belloni \& Hasinger}{1990a}]{belloni:90a}
Belloni, T., \& Hasinger, G.,  1990a, A\&A, 227, L33

\bibitem[\protect\astroncite{Belloni \& Hasinger}{1990b}]{belloni:90b}
Belloni, T., \& Hasinger, G.,  1990b, A\&A, 230, 230

\bibitem[\protect\astroncite{Bendat \& Piersol}{1986}]{bendat}
Bendat, J., \& Piersol, A.,  1986,
\newblock Random Data: Analysis and Measurement Procedures,
\newblock  (New York: Wiley)

\bibitem[\protect\astroncite{{Bochkar\"{e}v} et~al.}{1986}]{bochkarev:86a}
{Bochkar\"{e}v}, N.~G., Karitskaya, E.~A., Luskutov, V.~M., \& Sokolov, V.~V.,
  1986, Sov. Astron., 30, 43 (orig.: AZh, 63, 71, 1986)

\bibitem[\protect\astroncite{Bowyer et~al.}{1965}]{bowyer:65a}
Bowyer, S., Byram, E.~T., Chubb, T.~A., \& Friedman, H.,  1965, Science, 147,
  394

\bibitem[\protect\astroncite{Brainerd \& Lamb}{1987}]{brainerd:87a}
Brainerd, J., \& Lamb, F.~K.,  1987, ApJ, 317, L33

\bibitem[\protect\astroncite{Bussard et~al.}{1988}]{bussard:88a}
Bussard, R.~W., Weisskopf, M.~C., Elsner, R.~F., \& Shibazaki, N.,  1988, ApJ,
  327, 284

\bibitem[\protect\astroncite{Coppi}{1992}]{coppi:92a}
Coppi, P.~S.,  1992, MNRAS, 258, 657

\bibitem[\protect\astroncite{Crary et~al.}{1998}]{crary:98a}
Crary, D.~J., Finger, M.~H., {van der Hooft}, C. K.~F., {van Paradijs}, J.,
  {van der Klis}, M., \& Lewin, W. H.~G.,  1998, ApJ,
\newblock submitted

\bibitem[\protect\astroncite{Cui et~al.}{1997a}]{cui:96a}
Cui, W., Heindl, W.~A., Rothschild, R.~E., Zhang, S.~N., Jahoda, K., \& Focke,
  W.,  1997a, ApJ, 474, L57

\bibitem[\protect\astroncite{Cui et~al.}{1997b}]{cui:97b}
Cui, W., Zhang, S.~N., Focke, W., \& Swank, J.~H.,  1997b, ApJ, 484, 383

\bibitem[\protect\astroncite{Dolan}{1992}]{dolan:92a}
Dolan, J.~F.,  1992, ApJ, 384, 249

\bibitem[\protect\astroncite{Dove, Wilms \& Begelman}{1997}]{dove:97a}
Dove, J.~B., Wilms, J., \& Begelman, M.~C.,  1997, ApJ, 487, 747

\bibitem[\protect\astroncite{Dove et~al.}{1997}]{dove:97b}
Dove, J.~B., Wilms, J., Maisack, M.~G., \& Begelman, M.~C.,  1997, ApJ, 487,
  759

\bibitem[\protect\astroncite{Dove et~al.}{1998}]{dove:98a}
Dove, J.~B., Wilms, J., Nowak, M.~A., Vaughan, B.~A., \& Begelman, M.~C.,
  1998, MNRAS,
\newblock in press ~(paper I)

\bibitem[\protect\astroncite{ESA}{1997}]{esa:97a}
ESA, (eds.) 1997,
\newblock The {H}ipparcos and {T}ycho Catalogues,
\newblock  ESA SP 1200,  (Noordwijk: ESA Publications Division)

\bibitem[\protect\astroncite{{Gierli\'nski} et~al.}{1997}]{gierlinski:97a}
{Gierli\'nski}, M., Zdziarski, A.~A., Done, C., Johnson, W.~N., Ebisawa, K.,
  Ueda, Y., Haardt, F., \& Phlips, B.~F.,  1997, MNRAS,
\newblock in press

\bibitem[\protect\astroncite{Gies \& Bolton}{1982}]{gies:82a}
Gies, D.~R., \& Bolton, C.~T.,  1982, ApJ, 260, 240

\bibitem[\protect\astroncite{Gies \& Bolton}{1986}]{gies:86a}
Gies, D.~R., \& Bolton, C.~T.,  1986, ApJ, 304, 304

\bibitem[\protect\astroncite{Haardt \& Maraschi}{1993}]{haardt:93a}
Haardt, F., \& Maraschi, L.,  1993, ApJ, 413, 507

\bibitem[\protect\astroncite{Herrero et~al.}{1995}]{herrero:95a}
Herrero, A., Kudritzki, R.~P., Gabler, R., Vilchez, J.~M., \& Gabler, A.,
  1995, A\&A, 297, 556

\bibitem[\protect\astroncite{Hjellming}{1973}]{hjellming:73a}
Hjellming, R.~M.,  1973, ApJ, 182, L29

\bibitem[\protect\astroncite{Hua \& Titarchuk}{1995}]{hua:95a}
Hua, X.-M., \& Titarchuk, L.,  1995, ApJ, 449, 188

\bibitem[\protect\astroncite{Hutchings}{1978}]{hutchings:78a}
Hutchings, J.~B.,  1978, ApJ, 226, 264

\bibitem[\protect\astroncite{Jahoda et~al.}{1996}]{jahoda:96b}
Jahoda, K., Swank, J.~H., Giles, A.~B., Stark, M.~J., Strohmayer, T., Zhang,
  W., \& Morgan, E.~H.,  1996,
\newblock in {EUV}, X-Ray, and Gamma-Ray Instrumentation for Astronomy {VII},
  ed. O.~H. Siegmund,  (Bellingham, WA: SPIE), 59

\bibitem[\protect\astroncite{Kazanas, Hua \& Titarchuk}{1997}]{kazanas:97a}
Kazanas, D., Hua, X.-M., \& Titarchuk, L.,  1997, ApJ, 480, 280

\bibitem[\protect\astroncite{Kouveliotou et~al.}{1992}]{kouveliotou:92a}
Kouveliotou, C., Finger, M.~H., Fishman, G.~J., Meegan, C.~A., Wilson, R.~B.,
  \& Paciesas, W.~S.,  1992,
\newblock IAU Circ.~5576

\bibitem[\protect\astroncite{Kylafis \& Klimis}{1987}]{kylafis:87a}
Kylafis, N.~D., \& Klimis, G.~S.,  1987, ApJ, 323, 678

\bibitem[\protect\astroncite{Kylafis \& Phinney}{1989}]{kylafis:89a}
Kylafis, N.~D., \& Phinney, E.~S.,  1989,
\newblock in \cite{oegelman:89a}, 731

\bibitem[\protect\astroncite{Leahy et~al.}{1983}]{leahy:83a}
Leahy, D.~A., Darbro, W., Elsner, R.~F., Weisskopf, M.~C., Sutherland, P.~G.,
  Kahn, S., \& Grindlay, J.,  1983, ApJ, 266, 160

\bibitem[\protect\astroncite{Liang \& Nolan}{1984}]{liang:84a}
Liang, E.~P., \& Nolan, P.~L.,  1984, Space Sci. Rev., 38, 353

\bibitem[\protect\astroncite{Lomb}{1976}]{lomb:76a}
Lomb, N.~R.,  1976, Ap\&SS, 39, 447

\bibitem[\protect\astroncite{Lutz \& Lutz}{1972}]{lutz:72a}
Lutz, J.~H., \& Lutz, T.~E.,  1972, AJ, 77, 376

\bibitem[\protect\astroncite{Miller \& Lamb}{1992}]{miller:92a}
Miller, G.~S., \& Lamb, F.~K.,  1992, ApJ, 388, 541

\bibitem[\protect\astroncite{Miller}{1995}]{miller:95a}
Miller, M.~C.,  1995, ApJ, 441, 770

\bibitem[\protect\astroncite{Miyamoto et~al.}{1991}]{miyamoto:91a}
Miyamoto, S., Kimura, K., Kitamoto, S., Dotani, T., \& Ebisawa, K.,  1991, ApJ,
  383, 784

\bibitem[\protect\astroncite{Miyamoto \& Kitamoto}{1989}]{miyamoto:89a}
Miyamoto, S., \& Kitamoto, S.,  1989, Nature, 342, 773

\bibitem[\protect\astroncite{Miyamoto et~al.}{1992}]{miyamoto:92a}
Miyamoto, S., Kitamoto, S., Iga, S., Negoro, H., \& Terada, K.,  1992, ApJ,
  391, L21

\bibitem[\protect\astroncite{Morgan, Remillard \& Greiner}{1997}]{morgan:97a}
Morgan, E.~H., Remillard, R.~A., \& Greiner, J.,  1997, ApJ, 482, 993

\bibitem[\protect\astroncite{Narayan}{1996}]{narayan:96e}
Narayan, R.,  1996, ApJ, 462, 136

\bibitem[\protect\astroncite{{NASA}}{1997}]{nasa:97a}
{NASA},  1997,
\newblock {R}ossi {X}-ray Timing Explorer Guest Observer Program, Cycle 3,
\newblock Technical Report NRA 97-OSS-09,  (Washington, DC: NASA Office of
  Space Science)

\bibitem[\protect\astroncite{Ninkov, Walker \& Yang}{1987a}]{ninkov:87b}
Ninkov, Z., Walker, G. A.~H., \& Yang, S.,  1987a, ApJ, 321, 438

\bibitem[\protect\astroncite{Ninkov, Walker \& Yang}{1987b}]{ninkov:87a}
Ninkov, Z., Walker, G. A.~H., \& Yang, S.,  1987b, ApJ, 321, 425

\bibitem[\protect\astroncite{Nowak \& Vaughan}{1996}]{nowak:96a}
Nowak, M.~A., \& Vaughan, B.~A.,  1996, MNRAS, 280, 227

\bibitem[\protect\astroncite{Nowak et~al.}{1998}]{nowak:98b}
Nowak, M.~A., Wilms, J., Vaughan, B.~A., Dove, J., \& Begelman, M.~C.,  1998,
  ApJ,
\newblock submitted ~(paper III)

\bibitem[\protect\astroncite{Oda}{1977}]{oda:77a}
Oda, M.,  1977, Space Sci. Rev., 20, 757

\bibitem[\protect\astroncite{{\"Ogelman} \& {van den
  Heuvel}}{1989}]{oegelman:89a}
{\"Ogelman}, H., \& {van den Heuvel}, E. P.~J., (eds.) 1989,
\newblock Timing Neutron Stars,  NATO ASI C262,  (Dordrecht: Kluwer)

\bibitem[\protect\astroncite{Pottschmidt}{1997}]{pottschmidt:97a}
Pottschmidt, K.,  1997,
\newblock {D}i\-plom\-ar\-beit, {Eber\-hard-Karls-Uni\-ver\-si\-t\"at},
  T\"ubingen

\bibitem[\protect\astroncite{Pottschmidt et~al.}{1998}]{pottschmidt:98a}
Pottschmidt, K., {K\"onig}, M., Wilms, J., \& Staubert, R.,  1998, A\&A,
\newblock in press

\bibitem[\protect\astroncite{Poutanen, Krolik \& Ryde}{1997}]{poutanen:97b}
Poutanen, J., Krolik, J.~H., \& Ryde, F.,  1997, MNRAS, 221, 21p

\bibitem[\protect\astroncite{Poutanen, Svensson \& Stern}{1997}]{poutanen:97a}
Poutanen, J., Svensson, R., \& Stern, B.,  1997,
\newblock in \cite{winkler:97a}, 401

\bibitem[\protect\astroncite{Pozdnyakov, Sobol \& Sunyaev}{1983}]{pozd:83a}
Pozdnyakov, L.~A., Sobol, I.~M., \& Sunyaev, R.~A.,  1983, 2, 189

\bibitem[\protect\astroncite{Scargle}{1982}]{scargle:82a}
Scargle, J.~D.,  1982, ApJ, 263, 835

\bibitem[\protect\astroncite{Shakura \& Sunyaev}{1973}]{shakura:73a}
Shakura, N.~I., \& Sunyaev, R.,  1973, A\&A, 24, 337

\bibitem[\protect\astroncite{Shakura \& Sunyaev}{1976}]{shakura:76a}
Shakura, N.~I., \& Sunyaev, R.~A.,  1976, MNRAS, 175, 613

\bibitem[\protect\astroncite{Sokolov}{1987}]{sokolov:87a}
Sokolov, V.~V.,  1987, Sov. Astron., 31, 419 (orig.: AZh, 64, 64, 1987)

\bibitem[\protect\astroncite{Stollman et~al.}{1987}]{stollman:87a}
Stollman, G.~M., Hasinger, G., Lewin, W. H.~G., {van der Klis}, M., \& {van
  Paradijs}, J.,  1987, MNRAS, 227, 7p

\bibitem[\protect\astroncite{Sunyaev \& Titarchuk}{1980}]{st80}
Sunyaev, R.~A., \& Titarchuk, L.~G.,  1980, A\&A, 86, 121

\bibitem[\protect\astroncite{Sunyaev \& {Tr\"umper}}{1979}]{sunyaev:79a}
Sunyaev, R.~A., \& {Tr\"umper}, J.,  1979, Nature, 279, 506

\bibitem[\protect\astroncite{Svensson \& Zdziarski}{1994}]{svensson:94a}
Svensson, R., \& Zdziarski, A.~A.,  1994, ApJ, 436, 599

\bibitem[\protect\astroncite{Tanaka \& Lewin}{1995}]{tanaka:95a}
Tanaka, Y., \& Lewin, W. H.~G.,  1995,
\newblock in {X}-Ray Binaries, ed. W.~H.~G. Lewin, J. {van Paradijs}, E.~P.~J.
  {van den Heuvel},
\newblock  (Cambridge), Chapt.~3,  126

\bibitem[\protect\astroncite{Turon et~al.}{1992}]{turon:92a}
Turon, C., {Cr\'ez\'e}, M., Egret, D., {G\'omez}, A., et~al., 1992,
\newblock The Hipparcos Input Catalogue,
\newblock  ESA-SP 1136,  (Noordwijk: ESA Publications Division)

\bibitem[\protect\astroncite{Ubertini et~al.}{1994}]{ubertini:94a}
Ubertini, P., Bazzano, A., Cochhi, M., {L}a {P}udla, C., Polcaro, V.~F.,
  Staubert, R., \& Kendziorra, E.,  1994, ApJ, 421, 269

\bibitem[\protect\astroncite{{van der Klis}}{1989}]{vanderklis:89b}
{van der Klis}, M.,  1989,
\newblock in \cite{oegelman:89a}, 27

\bibitem[\protect\astroncite{Vaughan}{1991}]{vaughan:91a}
Vaughan, B.~A.,  1991,
\newblock Dissertation, Stanford University, Stanford, CA

\bibitem[\protect\astroncite{Vaughan \& Nowak}{1997}]{vaughan:97a}
Vaughan, B.~A., \& Nowak, M.~A.,  1997, ApJ, 474, L43

\bibitem[\protect\astroncite{Vikhlinin et~al.}{1994}]{vikhlinin:94a}
Vikhlinin, A., et~al., 1994, ApJ, 424, 395

\bibitem[\protect\astroncite{Walborn}{1973}]{walborn:73a}
Walborn, N.~R.,  1973, ApJ, 179, L123

\bibitem[\protect\astroncite{Wijers, {van Paradijs} \&
  Lewin}{1987}]{wijers:87a}
Wijers, R. A. M.~J., {van Paradijs}, J., \& Lewin, W. H.~G.,  1987, MNRAS, 228,
  17p

\bibitem[\protect\astroncite{Wilms et~al.}{1997}]{wilms:97a}
Wilms, J., Dove, J., Staubert, R., \& Begelman, M.~C.,  1997,
\newblock in \cite{winkler:97a}, 233

\bibitem[\protect\astroncite{Winkler, Courvoisier \&
  Durouchoux}{1997}]{winkler:97a}
Winkler, C., Courvoisier, T. J.-L., \& Durouchoux, P., (eds.) 1997,
\newblock The Transparent Universe,  ESA SP 382,  (Noordwijk: ESA Publications
  Division)

\bibitem[\protect\astroncite{Wu et~al.}{1982}]{wu:82a}
Wu, C.-C., Eaton, J.~A., Holm, A.~V., Milgrom, M., \&
  {Ham\-mer\-schlag-Hens\-ber\-ge}, G.,  1982, PASP, 94, 149

\bibitem[\protect\astroncite{Zhang et~al.}{1997}]{zhang:97b}
Zhang, S.~N., Cui, W., Harmon, B.~A., Paciesas, W.~S., Remillard, R.~E., \&
  {van Paradijs}, J.,  1997, ApJ,
\newblock in press

\bibitem[\protect\astroncite{Zhang \& Jahoda}{1996}]{zhangw:96a}
Zhang, W., \& Jahoda, K.,  1996,
\newblock Deadtime Effects in the {PCA},
\newblock Technical report,  (Greenbelt: NASA GSFC),
\newblock version dated 1996 September 26

\bibitem[\protect\astroncite{Zhang et~al.}{1995}]{zhangw:95a}
Zhang, W., Jahoda, K., Swank, J.~H., Morgan, E.~H., \& Giles, A.~B.,  1995,
  ApJ, 449, 930

\end{thebibliography}

\end{document}